\documentclass[referee]{aa}  
\usepackage{graphicx}
\usepackage{xcolor}
\usepackage{txfonts}
\begin{document}

   \title{Chemical and radiative transfer modeling of Propylene Oxide}

\titlerunning{Modeling of Propylene Oxide}

   \author{Ankan Das
          \inst{1}\fnmsep\thanks{ankan.das@gmail.com},
          Prasanta Gorai\inst{1}
          \and
          Sandip K. Chakrabarti\inst{1}}

   \institute{Indian Centre for Space Physics, 43 Chalantika, Garia Station Road, Kolkata 700084, India\\
              \email{ankan.das@gmail.com}
                                             }
\authorrunning{Das et al.}


  \abstract
{The recent identification of the first complex chiral molecule, propylene oxide (PrO) in space 
opens up a new window to further study the origin of homochirality on the Earth. 
{ There are some recent studies to explain the formation of PrO however
additional studies on the formation of this species are needed for better understanding.}}
{We seek to prepare a complete reaction network to study the formation of propylene oxide
in the astrophysically relevant conditions. 
Based on our results, a detailed radiative transfer modeling has been carried out 
to propose some more transitions which would potentially be targeted in the millimeter wave domain.}
{Gas-grain chemical network was used to explain the observed abundance of PrO in a cold shell surrounding the 
high-mass star-forming region of Sgr B2. Quantum chemical calculations were employed to study various reaction 
parameters and to compute multiple vibrational frequencies of PrO.}
{{ To model the formation of PrO in the observed region, 
we considered a dark cloud model.} Additionally, we used a model to check the feasibility of forming PrO in the hot core region. Some potential transitions in the millimeter wave domain are predicted which could be useful for the future astronomical detection. 
Radiative transfer modeling has been utilized to extract the physical condition
which might be useful to know the properties of the source in detail. 
Moreover, vibrational transitions of PrO has been provided which could be
very useful for the future detection of PrO by the upcoming
{ James Webb Space Telescope (JWST).}}
{} 

   \maketitle
%

\section{Introduction}
The discovery of a large number of complex interstellar species significantly improved our understanding of the
chemical processes involved in the Interstellar
Medium (ISM). It is speculated that the building blocks of life (biomolecules) were formed in the ISM and were
delivered by the comets to the Earth in the later stages \citep{chyb90,raym04,hart11}. The life making primordial
organic compounds may also have been formed in the pre-solar nebula rather than on the early Earth
\citep{bail98,chak00a,chak00b,hunt04,holt05,buse06,herb09}. The prospect of extra-terrestrial origin
of biomolecules is a fascinating topic since many biomolecules are chiral. { Synthesis and detection of prebiotic molecules in the
ISM have been studied and discussed by various authors \citep{cunn07,das08a,gupt11,maju12,garr13,nuev14,chak15}}.
In between the living organisms on the Earth, amino acids and sugars are chiral. Interestingly,
most of the amino acids of terrestrial proteins and sugars found on the earth are homochiral;
Amino acids are left-handed whereas sugars
are right-handed. When and how this homochirality developed on the Earth is a matter of debate \citep{cohe95}.
In the Murchison meteorite, various bio-molecules were identified. Interestingly,
an excess amount of L-amino acids were identified \citep{enge82,enge97} which also suggest an extraterrestrial
source for molecular asymmetry in the Solar System. Thus, observation of more chiral species along with their
enantiomeric excess in space could be very useful for the better understanding of the origin of
homochirality. But unfortunately, with the present observational facility, it is quite difficult to
define the enantiomeric excess of an interstellar species.

\cite{cunn07} attempted to observe PrO in the Orion KL and
Sgr B2(LMH) by using the Mopra Telescope. Based on their observational results, they
predicted an upper limit on the column density ($\sim 6.7 \times 10^{14}$ cm$^{-2}$) of PrO 
in the Sgr B2 (LMH) and predicted an excitation temperature of $200$ K with a compact source size of $5''$.
Recently, \cite{mcgu16} attempted to search for the existence of the
complex chiral species in space. They targeted the star-forming region like, Sgr B2 with Parkes and
Green Bank Telescope (GBT) and successfully identified the first complex chiral molecule (PrO) in space. 
They were unable to determine the enantiomeric excess of PrO
due to the lack of high precision and full polarization state measurement. \cite{mcgu16}
identified three transitions of PrO in absorption which correspond to the excitation temperature
of $\sim 5$ K, and column density $\sim1\times 10^{13}$ cm$^{-2}$.

PrO has a significant proton affinity \citep{hunt98} and should be converted into protonated form through
some ion-neutral reactions. Figure 1abc depict the structure of PrO and its two protonated form respectively.
\cite{swal57} and \cite{hers58} discussed the internal barrier of PrO and its various
rotational transitions in the ground and torsionally excited states. Later, \cite{cres77} calculated the centrifugal
distortional constants and structural parameters of PrO. The absorption spectrum
of PrO was studied by various authors \citep{pola85,lowe86,gont14}. Recently, the IR
spectrum of solid PrO was experimentally obtained by \cite{huds17}. { They also proposed
that the reaction between O atom and $\rm{C_3H_6}$ may lead to the formation of
PrO and its four other isomers. We have considered these pathways in our model to study the
formation of PrO and its isomers.}
\begin{figure}
{\centering
\includegraphics[width=9cm]{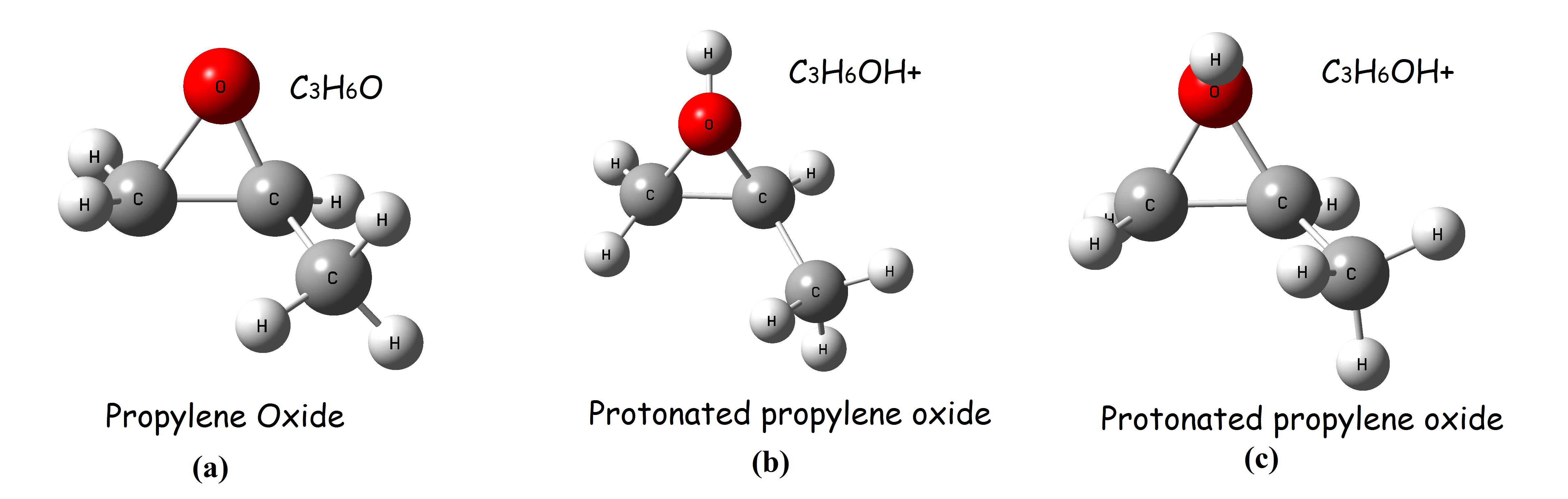}
\caption{Structure of propylene oxide and two of its protonated form.}}
\end{figure}

A Large number of observational and theoretical studies reveal that the complex organic molecules (COMs)
begin to form in the cold region of the dense molecular cloud.
{ Cosmic dust particles act as a reservoir for molecules and also as a
catalyst for reactions, which lead to the formation of COMs in molecular clouds
\citep{ober09,bacm12,das16}. }
{In low temperature and moderately high-density regions, gas phase exothermic reactions 
(such as ion-neutral reactions) are more feasible because they do not possess any activation barrier 
\citep{agun13}. At the same time, various irradiation-triggered solid-phase reactions also contribute to 
the formation of various complex interstellar species. In the low temperature, non-thermal desorption 
mechanism plays an efficient means to transport the surface species into the gas phase upon its formation. 
In this process, all the surface reactions which are exothermic and resulting in a single 
product can break the surface-molecule bond with some fraction and transfer the surface species 
to the gas phase \citep{garr07}. Non-thermal reactions on the grain surface also play a 
crucial role in governing the abundances of the surface species where the diffusion of atom (ground state) or 
radical diffusion chemistry is inefficient. Oxygen(${^1}$D) insertion reaction mentioned in \cite{berg17, berg18} 
can be characterized as the non-thermal reactions. These excited oxygens or supra-thermal oxygens(${^1}$D) are mainly 
produced  by the effects of secondary electrons generated in the path of cosmic rays when it is in contact 
with the grain mantle. Main source of these oxygens are the various oxygen rich species like, $\rm{CO_2}$, $\rm{H_2O}$ 
and CO.}

{
\cite{occh12} carried out theoretical modeling for the formation of Ethylene Oxide (EO) in space.
They were able to reproduce the observed abundance of EO successfully.
Additionally, they considered the reaction between $\rm{C_3H_6}$ and O on the surface
to include the formation of PrO. However, they did not include a complete reaction network
for the creation and destruction of PrO and its associated species.
More importantly, they did not consider the formation of $\rm{C_3H_6}$ on grain surface. Recently
\cite{hick16} provided some pathways for the formation of $\rm{C_3H_6}$ on the grain surface which eventually 
can produce PrO.}

In this paper, we study the synthesis of PrO along with two of its structural isomers, namely, propanal
and acetone in star-forming conditions. The radiative
transfer models have been carried out to identify the most active transition of PrO in
the millimeter(mm) wave regions. Absorption features of PrO and integral
absorption coefficients obtained with high-level quantum chemical methods could be handy
for its future detection by the high-resolution Stratospheric Observatory for Infrared Astronomy (SOFIA) and the most awaited JWST. The remainder of
this paper is as follows: Section 2 describes the reaction pathways considered in our modeling. 
In Section 3, we discuss chemical modeling. In Section 4, vibrational spectroscopy is presented.
Radiative transfer modeling is discussed in Section 5, and finally, in section 6 we draw our conclusions.

\begin{table}
\scriptsize
\centering
\caption{Surface Reactions For Propylene and Propylene Oxide Synthesis.}
\begin{tabular}{c|c|c}
\hline
{\bf Reaction }&{\bf Reaction}&{\bf E$_a$ in}\\
{\bf number(type) }&&{\bf Kelvin}\\
\hline
R1(NR)$^a$&$\rm{H + C_3 \rightarrow l-C_3H}$&0\\
&$\rm{\rightarrow }c-C_3H$&0\\
R2(RR)$^a$&$\rm{H + l-C_3H \rightarrow l-C_3H_2}$&0\\
&$\rm{\rightarrow c-C_3H_2}$&0\\
R3(RR)$^a$&$\rm{H + c-C_3H \rightarrow c-C_3H_2}$&0\\
&$\rm{\rightarrow l-C_3H_2}$&0\\
R4(NR)$^a$&$\rm{H + c,l-C_3H_2 \rightarrow C_3H_3 (CH_2CCH)}$&0\\
R5(RR)$^a$&$\rm{H + C_3H_3 \rightarrow CH_3CCH}$&0\\
&$\rm{\rightarrow CH_2CCH_2}$&0\\
R6(NR)$^a$&$\rm{H + CH_3CCH \rightarrow C_3H_5}$&2013\\
R7(NR)$^b$&$\rm{H + CH_2CCH_2 \rightarrow C_3H_5}$&0\\
R8(RR)$^a$&$\rm{H + C_3H_5 \rightarrow C_3H_6}$&0\\
R9(NR)$^a$&$\rm{H + C_3H_6 \rightarrow C_3H_7}$&1600\\
R10(RR)$^a$&$\rm{H + C_3H_7 \rightarrow C_3H_8}$&0\\
R11(NR)$^a$&$\rm{H + C_3H_8 \rightarrow C_3H_7+H_2}$&4000\\
R12(NR)&$\rm{O(^{3}P) + C_3H_6 \rightarrow CH_3COCH_3}$&28.6$^c$\\
R13(NR)&$\rm{O(^{3}P) + C_3H_6 \rightarrow CH_3CH_2CHO}$&30.6$^c$\\
R14(NR)&$\rm{O(^{3}P) + C_3H_6 \rightarrow 1-methyl\ vinyl\ alcohol}$&32.8$^c$\\
R15(NR)&$\rm{O(^{3}P) + C_3H_6 \rightarrow 2-methyl\ vinyl\ alcohol}$&33.8$^c$\\
R16(NR)&$\rm{O(^{3}P) + C_3H_6 \rightarrow PrO}$&40$^d$\\
R17(NR)&$\rm{ O(^{1}D)/O^* + C_3H_6 \rightarrow PrO}$&$^e$\\
R18(NR)&$\rm{CH + C_2H_6 \rightarrow C_3H_6+H}$&branching ratio 0.25$^e$\\
R19(NR)&$\rm{CH + C_2H_6 \rightarrow C_2H_4+CH_3}$&branching ratio 0.75$^e$\\
\hline
\end{tabular}\\
$^a$\cite{hick16}, $^b$\url{(http://kinetics.nist.gov/kinetics/)}, $^c$This work, $^d$\cite{ward11},
$^e$\cite{berg18}  
\end{table}

\begin{table}
\centering
\scriptsize{
\caption{Kinetics of the reaction between $\rm{C_3H_6}$ and $\rm{O}$.}
\begin{tabular}
{p{0.38in}|p{1.17in}|p{0.75in}|p{0.75in}}
\hline
{\bf Reaction }&{\bf Reaction}&\multicolumn{2}{|c}{\bf{Enthalpy of the reaction (Kcal/mol)}}\\
{\bf number (type) }&& {\bf Gas phase} &{\bf Ice phase} \\
&&{\bf(scaled $E_a$ in K)}&{\bf(scaled $E_a$ in K)}\\
&& {\bf (rate @30K in cm$^3$ s$^{-1}$)}&{\bf (rate @25K in $s^{-1}$)}\\
\hline
R12(NR)&$\rm{O(^{3}P) + C_3H_6 \rightarrow CH_3COCH_3}$&-112.87 (-17.4) ($9.7 \times 10^{-14}$)&-116.17 (28.6) ($2.0 \times 10^{-12}$)\\
R13(NR)&$\rm{O(^{3}P) + C_3H_6 \rightarrow CH_3CH_2CHO}$&-105.88 (-16.3) ($9.4 \times 10^{-14}$)&-108.51 (30.6) ($1.6 \times 10^{-12}$)\\
R14(NR)&$\rm{O(^{3}P) + C_3H_6 \rightarrow 1-MVA}$&-98.30 (-15.2) ($9.0 \times 10^{-14}$)&-101.25 (32.8) ($1.2 \times 10^{-12}$)\\
R15(NR)&$\rm{O(^{3}P) + C_3H_6 \rightarrow 2-MVA}$&-95.64 (-14.7) ($8.9 \times 10^{-14}$)&-98.05 (33.8) ($1.1 \times 10^{-12}$)\\
R16(NR)&$\rm{O(^{3}P) + C_3H_6 \rightarrow PrO}$&-81.09 (-12.5) ($8.3 \times 10^{-14}$)&-82.92 (40.0) ($5.2 \times 10^{-13}$)\\
\hline
\end{tabular}}
\end{table}

\section{Reaction pathways}

\subsection{Methodology}
To study the formation and destruction reactions of propylene oxide and associated species, we have 
considered both gas phase and grain surface reactions. To study the reaction pathways, we 
have employed quantum chemical methods.  We have carried out quantum chemical calculations 
to find out the reaction enthalpies of these gas/ice phase reactions. 
In Table 1, we have summarized all the surface reactions which are considered for the formation of
propylene and propyline oxide.
In Table 2, we have provided the enthalpy of reactions between ${\rm C_3H_6}$ 
and O($^3$P) for five (R12-R16) product channels. Gaussian 09 program \citep{fris15} has been utilized for such calculations. 
Here, we have used Density Functional Theorem (DFT) with 6-311G basis set with the inclusion of diffuse (++) along with polarization functions 
(d,p) and B3LYP functional \citep{beck88,lee88}. The Polarizable Continuum Model (PCM) using the integral equation formalism variant (IEFPCM) with 
the default SCRF method has been used to model the solvation-effect for the computation of enthalpy of the ice phase reactions.

\subsection{Formation pathways}
\cite{marc07} had discovered propylene ($\rm{C_3H_6}$) towards the TMC-1 by using the IRAM $30$m radio
telescope. They estimated the column density of propylene $\sim {4\times 10^{13}}$ $\rm{cm^{-2}}$. 
\cite{herb10} studied the formation of gas phase propylene in the cold interstellar cloud. 
They proposed that propylene could mainly be formed by the dissociative recombination of its 
precursor ion, $\rm{H_3CCHCH_3^{+}}$ via
$
\rm{C_3H_7^{+} + e^{-}\rightarrow C_3H_6}+H.
$
For the formation of $\rm{C_3H_7^{+}}$, 
following reactions were included in UMIST 2012 \citep{mcel13} network:
$$
\rm{{C_3H_3^{+}}+H_2 \rightarrow C_3H_5^{+}},
$$
$$
\rm{{C_3H_5^{+}}+H_2 \rightarrow C_3H_7^{+}}.
$$
But recent study by \cite{lin13} found that above two reactions contain some activation barrier ($25.5$ and $24.7$ kJ/mol respectively). 
In the low-temperature regime, it is problematic to overcome such a high barrier in the gas phase and thus 
the formation of gas phase $\rm{C_3H_7^{+}}$ is inadequate by these routes which suggest considering some alternative
reaction pathways to enable the formation of $\rm{C_3H_6}$ rather than by 
the electron recombination reaction as proposed by \cite{herb10}.
Recently, \cite{hick16} proposed some pathways for the formation of $\rm{C_3H_6}$ 
in the ice phase. Here also, we have considered their 
pathways too for the formation of this species.
In Table 1, we have listed the formation and destruction pathways of $\rm{C_3H_6}$ found in the literature and calculated in this work. Though we have considered the formation of $\rm{C_3}$ on the grain surface, around the 
low temperature, its formation on the grain surface is not adequate due to the higher binding energy 
(~$4000$K from the KIDA database) of the C atom. 
$\rm{C_3}$ is efficiently forming on the gas phase, and thus the formation of $\rm{C_3H_6}$ begins after 
the accretion of gas phase $\rm{C_3}$ on the grain surface. 
Sequential hydrogenation reactions (R1-R8 of Table 1) in the ice phase then leads to the formation of $\rm{C_3H_6}$.
Most of these hydrogenation reactions are barrier-less (only reaction R6 has some barrier), so these pathways
are very efficient for the formation of $\rm{C_3H_6}$ in ice phase. 
 Recently, \cite{berg18} used the reaction between CH and $\rm{C_2H_6}$ for the formation of propylene with
a branching fraction of $0.25$ (R18). We also have considered reaction number R18 and R19 ($0.75$ branching fraction) 
of Table 1 in our calculations as well.
A high abundance of ice phase species may reflect in the gas phase by various efficient desorption mechanisms 
such as thermal, non-thermal and cosmic-ray induced desorption.

The gas phase reaction between oxygen atom O($^{3}$P) and propylene ($\rm{C_3H_6}$) has been well studied by 
several authors \citep{stuh71,atki72,atki74,knya92}. 
Branching ratios of the reaction between 
$\rm{C_3H_6}$ and O atom were studied by \cite{debo07}. They showed that the propylene and oxygen atom could form a stable bi-radical ($\rm{CH_3CHCH_2O}$) having an activation barrier of $\sim7.5$ kJ/mol. Dissociation of this bi-radical follows various pathways with both singlet and triplet potential energy surfaces. Isomerization
of PrO can produce (a) acetone, (b) propanal (c) methyl vinyl alcohol and 
(d) ally alcohol. 
However, all these have to overcome very high activation barriers \citep{dubn00} which is not possible at our desired environment.\\

{ \cite{ward11} described the production of EO and PrO on interstellar dust (graphite surface). 
They found an activation barrier of about $40$ K for the formation of PrO by the reaction between
$\rm{C_3H_6}$ and O($^3$P).}
\cite{huds17} proposed that the reaction between O($^3$P) atom and $\rm{C_3H_6}$ can produce any of 
the following five species, namely, (i) propylene oxide (${\rm C_3H_6O}$), (ii) propanol (${\rm CH_3CH_2CHO}$), 
(iii) 1-methyl vinyl alcohol (${\rm CH_3COHCH2}$),
(iv) acetone (${\rm CH_3COCH_3}$) and (v) 2-methyl vinyl alcohol (${\rm CH_3CHCHOH}$). 
{ We have carried out quantum chemical calculations to find out the reaction enthalpies of these gas/ice phase reactions. 
In Table 2, we have shown the enthalpy of reactions between ${\rm C_3H_6}$ and O($^3$P)
for these five (R12-R16) product channels.}
Table 2 depicts that in between the five product channels (R12-R16), 
production of acetone is comparatively most, and production of PrO is least favorable.
In Table 1, reaction R12-R16 are arranged according to their exothermicity values obtained from our calculated values in Table 2. \cite{ward11} found an activation barrier of about $40$ K for the formation of ice phase PrO. 
We have considered the activation barrier of the other product channels by using a scaling factor 
based on their exothermicity values obtained from the ice phase reactions of Table 2.
The ice phase rate coefficients of these reactions (R12-R16) { have been} calculated by the method described in \cite{hase92}, which is based on thermal diffusion. 
For the binding energy of $\rm{O}$, we have used $1660$ K \citep{he15} and for $\rm{C_3H_6}$, { we have used 
$2580$ K \citep{ward11}} respectively. 
For the other species, we have considered the most updated set of energy barriers available in KIDA 
database. 
Calculated ice phase rate for these $5$ product channels are shown in the last column of Table 2 for ($T_{ice}=25$ K).

Recently \cite{leno15} computed the rate constant for the formation of gas phase PrO at high temperature for
the addition and abstraction reactions interpolated between $300-2000$ K. 
According to their results, formation of PrO by the addition
of $\rm{C_3H_6}$ and O($^3$P) can be approximated by the following Arrhenius expression:
$$
K=\alpha T^\beta exp(-E_a/T) \ \ cm^3 \ molecules^{-1} \ s^{-1},
$$
where, $\alpha= 4.52 \times 10^{-16}$, $\beta=1.406$ and $E_a$, the activation barrier is $-12.5$ K. Here,
we have assumed that this reaction is also feasible around the low temperature as well. Using the 
above relation, at $30$ K, we have a rate of $8.3 \times 10^{-14}$ cm$^3$ s$^{-1}$  
for the formation of gas phase PrO (R16 of Table 2).
Since the activation barrier for the other $4$ products resulting from the gas phase reaction 
between $\rm{C_3H_6}$ and O (R12-R15 of Table 2) are yet to be known,
we have used their gas phase exothermicity values pointed out in Table 2 
to scale the activation barrier ($E_a$). 
Thus, for these $4$ product 
channels, we have kept $\alpha$ and $\beta$ same as the gas phase reaction number R16 of Table 2 and
have scaled $E_a$ according to their exothermicity values.


\cite{dick97} proposed the following gas-phase formation pathways of EO:
$$\rm{CH_3^{+}+ C_2H_5OH \rightarrow C_2H_5O^{+} + CH_4}$$ followed by,
$$\rm{C_2H_5O^{+} + e^{-} \rightarrow C_2H_4O+H}.$$
\clearpage
\onecolumn
\begin{table}
\centering
\scriptsize{
\caption{Destruction pathways of PrO and its isomers.}
\begin{tabular}{|p{1.3cm}|c|c|c|c|c|}
\hline
\hline
{\bf Reaction number (type)} & {\bf Reaction} & $\alpha$ & $\beta$ & $\gamma$  & {\bf Rate coefficient@$30$ K} \\
\hline
R1 (IN) & $\mathrm{C^+ + C_3H_6O\rightarrow C_3H_6O^+ + C}$ & $1.50\times10^{-9}$ & -0.5 & 0.0 & $4.74\times10^{-09}$ \\
R2 (IN) & $\mathrm{C^+ + CH_3CH_2CHO\rightarrow C_3H_6O^+ + C}$ & $1.50\times10^{-9}$ & -0.5 & 0.0 & $4.74\times10^{-09}$  \\
R3 (IN) & $\mathrm{C^+ + CH_3CHCHOH\rightarrow C_3H_6O^+ + C}$ & $1.50\times10^{-9}$ & -0.5 & 0.0 & $4.74\times10^{-09}$  \\
R4 (IN) &$\mathrm{C^+ + CH_3COHCH_2\rightarrow C_3H_6O^+ + C}$ & $1.50\times10^{-9}$ & -0.5 & 0.0 & $4.74\times10^{-09}$  \\
R5 (IN) & $\mathrm{C^+ + C_3H_6O\rightarrow C_3H_5O^+ + CH}$ & $1.50\times10^{-9}$ & -0.5 & 0.0 & $4.74\times10^{-09}$  \\
R6 (IN) & $\mathrm{C^+ + CH_3CH_2CHO\rightarrow C_3H_5O^+ + CH}$ & $1.50\times10^{-9}$ & -0.5 & 0.0 & $4.74\times10^{-09}$  \\
R7 (IN) & $\mathrm{C^+ + CH_3CHCHOH\rightarrow C_3H_5O^+ + CH}$ & $1.50\times10^{-9}$ & -0.5 & 0.0 & $4.74\times10^{-09}$  \\
R8 (IN) &$\mathrm{C^+ + CH_3COHCH_2\rightarrow C_3H_5O^+ + CH}$ & $1.50\times10^{-9}$ & -0.5 & 0.0 & $4.74\times10^{-09}$  \\
R9 (IN) & $\mathrm{H^+ + C_3H_6O\rightarrow C_3H_6O^+ + H}$ & $3.60\times10^{-9}$ & -0.5 & 0.0 & $1.13\times10^{-08}$ \\
R10 (IN) & $\mathrm{H^+ + CH_3CH_2CHO\rightarrow C_3H_6O^+ + H}$ & $3.60\times10^{-9}$ & -0.5 & 0.0 & $1.13\times10^{-08}$ \\
R11 (IN) & $\mathrm{H^+ + CH_3CHCHOH\rightarrow C_3H_6O^+ + H}$ & $3.60\times10^{-9}$ & -0.5 & 0.0 & $1.13\times10^{-08}$ \\
R12 (IN) &$\mathrm{H^+ + CH_3COHCH_2\rightarrow C_3H_6O^+ + H}$ & $3.60\times10^{-9}$ & -0.5 & 0.0 & $1.13\times10^{-08}$ \\
R13 (IN) & $\mathrm{H^+ + C_3H_6O\rightarrow C_3H_5O^+ + H}$ & $3.60\times10^{-9}$ & -0.5 & 0.0 & $1.13\times10^{-08}$ \\
R14 (IN) & $\mathrm{H^+ + CH_3CH_2CHO\rightarrow C_3H_5O^+ + H}$ & $3.60\times10^{-9}$ & -0.5 & 0.0 & $1.13\times10^{-08}$\\
R15 (IN) & $\mathrm{H^+ + CH_3CHCHOH\rightarrow C_3H_5O^+ + H}$ & $3.60\times10^{-9}$ & -0.5 & 0.0 & $1.13\times10^{-08}$ \\
R16 (IN) &$\mathrm{H^+ + CH_3COHCH_2\rightarrow C_3H_5O^+ + H}$ & $3.60\times10^{-9}$ & -0.5 & 0.0 & $1.13\times10^{-08}$ \\
R17 (IN) & $\mathrm{He^+ + C_3H_6O\rightarrow HCO + C_2H_5^+ + He}$ & $3.00\times10^{-9}$ & -0.5 & 0.0 & $9.48\times10^{-09}$ \\
R18 (IN) & $\mathrm{He^+ + CH_3CH_2CHO\rightarrow HCO + C_2H_5^+ + He}$ & $3.00\times10^{-9}$ & -0.5 & 0.0 & $9.48\times10^{-09}$ \\
R19 (IN) & $\mathrm{He^+ + CH_3CHCHOH\rightarrow HCO + C_2H_5^+ + He}$ & $3.00\times10^{-9}$ & -0.5 & 0.0 & $9.48\times10^{-09}$ \\
R20 (IN) &$\mathrm{He^+ + CH_3COHCH_2\rightarrow HCO + C_2H_5^+ + He}$ & $3.00\times10^{-9}$ & -0.5 & 0.0 & $9.48\times10^{-09}$ \\
R21 (IN) & $\mathrm{He^+ + C_3H_6O\rightarrow HCO^+ + C_2H_5 + He}$ & $3.00\times10^{-9}$ & -0.5 & 0.0 & $9.48\times10^{-09}$ \\
R22 (IN) & $\mathrm{He^+ + CH_3CH_2CHO\rightarrow HCO^+ + C_2H_5 + He}$ & $3.00\times10^{-9}$ & -0.5 & 0.0 &$9.48\times10^{-09}$ \\
R23 (IN) & $\mathrm{He^+ + CH_3CHCHOH\rightarrow HCO^+ + C_2H_5 + He}$ & $3.00\times10^{-9}$ & -0.5 & 0.0 & $9.48\times10^{-09}$ \\
R24 (IN) &$\mathrm{He^+ + CH_3COHCH_2\rightarrow HCO^+ + C_2H_5 + He}$ & $3.00\times10^{-9}$ & -0.5 & 0.0 & $9.48\times10^{-09}$ \\
R25 (IN) & $\mathrm{H_3^+ + C_3H_6O\rightarrow C_3H_7O^+ + H_2}$ & $4.14\times10^{-10}$ & -0.5 & 0.0 & $1.29\times10^{-09}$ \\
R26 (IN) & $\mathrm{H_3^+ + CH_3CH_2CHO\rightarrow C_3H_7O^+ + H_2}$ & $4.14\times10^{-10}$ & -0.5 & 0.0 & $1.29\times10^{-09}$ \\
R27 (IN) & $\mathrm{H_3^+ + CH_3CHCHOH\rightarrow C_3H_7O^+ + H_2}$ & $4.14\times10^{-10}$ & -0.5 & 0.0 & $1.29\times10^{-09}$ \\
R28 (IN) &$\mathrm{H_3^+ + CH_3COHCH_2\rightarrow C_3H_7O^+ + H_2}$ & $4.14\times10^{-10}$ & -0.5 & 0.0 & $1.29\times10^{-09}$ \\
R29 (IN) & $\mathrm{HCO^+ + C_3H_6O\rightarrow C_3H_7O^+ + CO}$ & $3.40\times10^{-9}$ & -0.5 & 0.0 & $1.07\times10^{-08}$ \\
R30 (IN) & $\mathrm{HCO^+ + CH_3CH_2CHO\rightarrow C_3H_7O^+ + CO}$ & $3.40\times10^{-9}$ & -0.5 & 0.0 & $1.07\times10^{-08}$ \\
R31 (IN) & $\mathrm{HCO^+ + CH_3CHCHOH\rightarrow C_3H_7O^+ + CO}$ & $3.40\times10^{-9}$ & -0.5 & 0.0 & $1.07\times10^{-08}$\\
R32 (IN) &$\mathrm{HCO^+ + CH_3COHCH_2\rightarrow C_3H_7O^+ + CO}$ & $3.40\times10^{-9}$ & -0.5 & 0.0 & $1.07\times10^{-08}$ \\
R33 (IN) & $\mathrm{H_3O^+ + C_3H_6O\rightarrow C_3H_7O^+ + H_2O}$ & $3.60\times10^{-9}$ & -0.5 & 0.0 & $1.13\times10^{-08}$ \\
R34 (IN) & $\mathrm{H_3O^+ + CH_3CH_2CHO\rightarrow C_3H_7O^+ + H_2O}$ & $3.60\times10^{-9}$ & -0.5 & 0.0 & $1.13\times10^{-08}$ \\
R35 (IN) & $\mathrm{H_3O^+ + CH_3CHCHOH\rightarrow C_3H_7O^+ + H_2O}$ & $3.60\times10^{-9}$ & -0.5 & 0.0 & $1.13\times10^{-08}$ \\
R36 (IN) &$\mathrm{H_3O^+ + CH_3COHCH_2\rightarrow C_3H_7O^+ + H_2O}$ & $3.60\times10^{-9}$ & -0.5 & 0.0 & $1.13\times10^{-08}$ \\

&&&&&\\
R37 (DR) & $\mathrm{C_3H_6O^+ + e^- \rightarrow C_2H_5 + HCO}$ & $1.08\times10^{-6}$ & -0.70 & 0.0 & $5.51\times10^{-06}$ \\
R38 (DR) & $\mathrm{C_3H_6O^+ + e^- \rightarrow CH_3CHCH_2 + O}$ & $1.08\times10^{-6}$ & -0.70 & 0.0 & $5.51\times10^{-06}$ \\
R39 (DR) & $\mathrm{C_3H_6O^+ + e^- \rightarrow CH_2CCH + H_2O + H}$ & $1.50\times10^{-7}$ & -0.50 & 0.0 & $4.74\times10^{-07}$ \\
R40 (DR) & $\mathrm{C_3H_6O^+ + e^- \rightarrow CH_3CCH + H_2O}$ & $1.50\times10^{-7}$ & -0.50 & 0.0 & $4.74\times10^{-07}$ \\
R41 (DR) & $\mathrm{C_3H_7O^+ + e^- \rightarrow C_2H_4 + H_2CO + H}$ & $8.47\times10^{-7}$ & -0.74 & 0.0 & $4.67\times10^{-06}$ \\
R42 (DR) & $\mathrm{C_3H_7O^+ + e^- \rightarrow C_2H_5 + HCO + H}$ & $8.47\times10^{-7}$ & -0.74 & 0.0 & $4.67\times10^{-06}$ \\
R43 (DR) & $\mathrm{C_3H_7O^+ + e^- \rightarrow CO + CH_3CH_3 + H}$ & $8.47\times10^{-7}$ & -0.74 & 0.0 & $4.67\times10^{-06}$ \\
R44 (DR) & $\mathrm{C_3H_7O^+ + e^- \rightarrow H_2CO + C_2H_5}$ & $8.47\times10^{-7}$ & -0.74 & 0.0 & $4.67\times10^{-06}$\\
R45 (DR) & $\mathrm{C_3H_7O^+ + e^- \rightarrow C_3H_6 + H}$ & $3.00\times10^{-7}$ & -0.74 & 0.0 & $1.64\times10^{-06}$ \\
R46 (DR) & $\mathrm{C_3H_7O^+ + e^- \rightarrow CH_3CH_2CHO + H}$ & $3.00\times10^{-7}$ & -0.74 & 0.0 & $1.64\times10^{-06}$ \\
R47 (DR) & $\mathrm{C_3H_7O^+ + e^- \rightarrow CH_3CHCHOH + H}$ & $3.00\times10^{-7}$ & -0.74 & 0.0 &$1.64\times10^{-06}$ \\
R48 (DR) & $\mathrm{C_3H_7O^+ + e^- \rightarrow CH_3COHCH_2 + H}$ & $3.00\times10^{-7}$ & -0.74 & 0.0 &$1.64\times10^{-06}$\\
R49 (DR) & $\mathrm{C_3H_5O^+ + e^- \rightarrow CH_3CHO + H}$ & $3.00\times10^{-7}$ & -0.74 & 0.0 & $9.48\times10^{-07}$ \\
R50 (DR) & $\mathrm{C_3H_5O^+ + e^- \rightarrow C_2H_5 + CO}$ & $3.00\times10^{-7}$ & -0.74 & 0.0 & $9.48\times10^{-07}$ \\
&&&&&\\
R51 (NR) & $\mathrm{CH_3CHCH_2 + O \rightarrow C_2H_5 + HCO}$ & $3.6\times10^{-12}$ & 0.0 & 0.0 & $3.60\times10^{-12}$ \\
R52 (NR) & $\mathrm{CH_3CHCH_2 + O \rightarrow CH_3CO + CH_3}$ & $4.4\times10^{-12}$ & 0.0 & 0.0 & $4.40\times10^{-12}$ \\
R53 (NR) & $\mathrm{CH_2CHCH_2 + N \rightarrow CH_2CHCN + H_2}$ & $3.2\times10^{-11}$ & 0.17 & 0.0 & $2.16\times10^{-11}$ \\
R54 (NR) & $\mathrm{CH_2CHCH_2 + N \rightarrow C_2H_4 + HCN}$ & $3.2\times10^{-11}$ & 0.17 & 0.0 & $2.16\times10^{-11}$ \\
R55 (NR) & $\mathrm{CH_2CHCH_2 + N \rightarrow C_3H_4 + NH}$ & $1.3\times10^{-11}$ & 0.17 & 0.0 & $8.78\times10^{-12}$ \\
R56 (NR) & $\mathrm{CH_3CHCH + N \rightarrow CH_2CHCN + H_2}$ & $3.2\times10^{-11}$ & 0.17 & 0.0 & $2.16\times10^{-11}$ \\
R57 (NR) & $\mathrm{CH_3CHCH + N \rightarrow C_2H_4 + HCN}$ & $3.2\times10^{-11}$ & 0.17 & 0.0 & $2.16\times10^{-11}$ \\
R58 (NR) & $\mathrm{CH_3CHCH + N \rightarrow C_3H_4 + NH}$ & $1.3\times10^{-11}$ & 0.17 & 0.0 & $8.78\times10^{-12}$ \\
R59 (PH) & $\mathrm{CH_3CHCH_2 + PHOTON \rightarrow C_2H_4 + CH_2}$ & $1.13\times10^{-09}$ & 0.00 & 1.6 & $1.56\times10^{-30}$ \\
R60 (CRPH) & $\mathrm{CH_3CHCH_2 + CR-PHOTON \rightarrow C_2H_4 + CH_2}$ & $1.30\times10^{-17}$ & 0.00 & 750 & $1.95\times10^{-14}$ \\
\hline
\hline
\end{tabular}}
\end{table}
\clearpage
\noindent Following the same trend as EO, here also, we have assumed that all the isomers of PrO could also be produced by the dissociative recombination of $\rm{C_3H_7O^+}$ with the same rate. Formation
of $\rm{C_3H_7O^+}$ ($\rm{CH_3COCH_4}^+$, i.e., protonated acetone or ethyl,
1- methoxy - , ion) was
already included in the UMIST 2012 network, so we have not used any additional channel to form this ion.
Instead, only the destruction of this ion by the dissociative recombination is considered.

\subsection{ Formation by supra-thermal oxygen}
Recently, \cite{berg18} carried out a combined theoretical and experimental work to study the formation of
interstellar PrO. They found that the galactic cosmic ray induced ice phase chemistry with the
supra-thermal oxygen (O$^*$ or O($^1$D)) may lead to the significant amount of PrO (R17 of Table 1).
For the generation of supra-thermal oxygen, they considered the dissociation of CO$_2$, H$_2$O and CO.
The rate constants of the radiolysis were calculated by using the following relation:
$$
R=\frac{\zeta}{10^{-17} [s^-1]} \frac{G}{100 [eV]}S_e \phi_{ism} \ \ \ s^{-1},
$$
where, G is the radiolysis yield per $100$eV, $S_e$ is the electronic stopping cross section,
$\Phi_{ism}$ is the interstellar cosmic ray flux and $\zeta$ is the cosmic ray ionization rate.
Here, we have used same $G$ value as used in \cite{berg18}
($2.23$, $0.7$ and $0.8$ for CO$_2$, H$_2$O and CO respectively), $\phi_{ism}=10$ cm$^{-2}$ s$^{-1}$ following
\cite{abpl16}. We have used $\zeta =1.3 \times 10^{-17}$ s$^{-1}$ instead of $1.3 \times 10^{-16}$ used 
in \cite{berg18}. Though \cite{abpl16} mentioned that $S_e$ is the
electron stopping cross section for electrons, \cite{berg18} used $S_e$ as the
electron stopping cross section for protons from the PSTAR programme.
Here, we have used ESTAR programme as used by \cite{abpl16}
(\url{https://physics.nist.gov/PhysRefData/Star/Text/ESTAR.html})
for the computation of the $S_e$ parameter.
Since, $S_e$ is sensitive to the projectile energy, for a better
approximation for $S_e$, we have used projectile energy of the electron
$1$ KeV (minimum available value in the ESTAR programme),
$5$ keV and $10$ KeV respectively and have considered their average.
We have found $S_e=3.62 \times 10^{-15}$, $1.80 \times 10^{-15}$ and $2.34 \times 10^{-15}$ ev cm$^2$/molecule when
we have used CO$_2$, H$_2$O and CO ice as target materials respectively. For the
density of the target materials, we have used $1.3$, $1$ and $0.81$ gm/cm$^3$ respectively for
CO$_2$,  H$_2$O and CO ice from \cite{lauc15} and references therein. By considering these $S_e$ parameters, we have obtained
rate constants of  $1.05 \times 10^{-13}$, $1.64 \times 10^{-14}$ and $2.43 \times 10^{-14}$ s$^{-1}$ respectively
for CO$_2$, H$_2$O and CO ice as a target material.


For the consideration of the supra-thermal oxygen (O($^1$D)) in our network, we have considered some trivial approximations.
After the generation of these supra-thermal oxygens, they are treated similarly as the O($^3$P), except
the reaction with ${\rm C_3H_6}$. For the reaction between $\rm{C_3H_6}$ and  O($^1$D), we have assumed that the
reaction is barrier-less in nature and process in each encounter. 
The binding energy of the O($^1$D) is assumed to be the same as O($^3$P)
and supposed to behave similarly in the gas phase upon its sublimation. 
We have not considered the O($^1$D) insertion reactions for any other species in our network. We have used it
only for the formation of PrO. We have
tested the effect of this oxygen insertion reaction on the abundances of the major
species which remain mostly unaltered.

\subsection{Destruction pathways}
All the destruction pathways for the PrO along with its isomers are shown in Table 3.
We have considered Ion-Neutral (IN) reactions for the destruction of PrO and its isomers. The most abundant 
ions like, $\rm{C^{+}, H^{+}, He^{+}, H_3^{+}, H_3O^{+},}$ and $\rm{HCO^{+}}$ have been considered 
\citep{occh12}. Additionally, dissociative recombination (DR), photo-dissociation (PH) and dissociation by cosmic rays (CRPH) 
have also been considered for the destruction.
Ion-Neutral reactions are very efficient for the destruction of neutral interstellar species. If the neutral species is non-polar, 
we have used the Langevin collision rate coefficient \citep{herb06,wake10} and if the neutral species is polar, then we have employed the trajectory scaling relation \citep{su82,woon09}. For the destruction by photo-reactions, 
we have used the following equation:
\begin{equation}
k_{PH} = \alpha \exp(-\gamma A_V )  ,
\end{equation}
where, $\alpha$ is the rate coefficient (s$^{-1}$) in the unshielded interstellar ultraviolet radiation field, $A_V$ is 
the visual extinction and $\gamma$ controls the extinction of the dust at the ultraviolet wavelength.
Here, we have used $\alpha = 1.13 \times 10^{-09}$ s$^{-1}$,  and $\gamma = 1.6$ for the photo-dissociation of $\rm{CH_3CHCH_2}$.\\

For the cosmic-ray-induced photo-reactions, we have used the following relation \citep{gred89}:
\begin{equation}
 k_{CRPH}=\alpha \gamma'/(1- \omega),
\end{equation}
where, $\alpha$ is the cosmic-ray ionization rate (s$^{-1}$), $\gamma'$ is the number of photo-dissociative events that 
take place per cosmic-ray ionization and $\omega$ is the dust grain albedo in the far ultraviolet. 
Here, we have used $\omega$ =0.6,  
$\alpha=1.3 \times 10^{-17}$ s$^{-1}$, and $\gamma'=750.0$  by following the cosmic ray induced 
photo-reactions of $\rm{CH_3OH}$ in \cite{wood07}.

%

\section{Chemical Modeling}
Here, we have used our large gas-grain chemical network \citep{das08b,das13a,das13b,maju14a,maju14b,das15a,das15b,
gora17a,gora17b} to estimate the formation of PrO in the star forming region. 
The gas phase chemical network is mainly adopted from the UMIST 2012 database \citep{mcel13}, and grain surface chemical
network is primarily developed from \cite{ruau16}. As the initial elemental abundances (Table 4), we have
considered the ``low metal" abundance having C/O ratio $\sim 0.607$ \citep{hinc11} which 
are often used for the modeling of dense clouds { where the majority of the hydrogen atoms are
locked in the form of hydrogen molecules.} 


\begin{table*}
\centering{
\scriptsize
\caption{Initial elemental abundances with respect to total hydrogen nuclei.}
\begin{tabular}{|c|c|}
\hline
Species&Abundance\\
\hline\hline
$\mathrm{H_2}$ &    $5.00 \times 10^{-01}$\\
$\mathrm{He}$  &    $0.90 \times 10^{-01}$\\
$\mathrm{N}$   &    $6.20 \times 10^{-05}$\\
$\mathrm{O}$   &    $2.80 \times 10^{-04}$\\
$\mathrm{C}$ &    $1.70 \times 10^{-04}$\\
$\mathrm{Na}$ &    $2.00 \times 10^{-09}$\\
$\mathrm{Mg}$ &    $7.00 \times 10^{-09}$\\
$\mathrm{Si}$ &    $8.00 \times 10^{-09}$\\
$\mathrm{P}$ &    $2.00 \times 10^{-10}$\\
$\mathrm{S}$ &    $8.00 \times 10^{-08}$\\
$\mathrm{Cl}$ &    $1.00 \times 10^{-09}$\\
$\mathrm{Fe}$ &    $3.00 \times 10^{-09}$\\
\hline
\end{tabular}}
\end{table*}

\subsection{Dark cloud model}
\subsubsection{Physical condition}
To mimic the observed region of the PrO, we have considered a total hydrogen number density 
($n_H$) $=2 \times 10^4$ cm$^{-3}$ and a gas temperature ($T_{gas})=30$ K and dust temperature ($T_{ice})=10$ and $25$ K respectively.
A visual extinction parameter ($A_V$) of $30$ have been assumed and a cosmic ray ionization rate ($\zeta$) of 
$1.3 \times 10^{-17}$ s$^{-1}$ \citep{herb73} been considered. 
Since the reactive desorption \citep{garr07} is a very efficient means to transfer surface species into the gas phase 
upon formation, we have considered it in our model.

\subsubsection{Results \& Discussions}
Chemical evolution of PrO and $\rm{C_3H_6}$ is shown in Fig. 2. The left panel of Fig. 2 is for
$T_{ice}=10$ K and right panel is for $T_{ice}=25$ K. Left panel depicts two scenarios, and one is with
a fiducial reactive desorption parameter $a=0.03$, unless otherwise stated, we always have used this value of desorption
factor along with the desorption probability $1$ {(for the reactions producing single product)} and another with $a=0$.
It shows that 
when reactive desorption parameter is absent ($a=0$), the formation of PrO is insignificant in both the phases. 
The reason behind is
that $\rm{C_3H_6}$ can efficiently be formed on ice phase but when $a=0.03$ is considered, significant portion 
of $\rm{C_3H_6}$ can be able to transfer to the gas phase which is not possible for $a=0$.
\cite{wood07} considered barrier-less reactions for the formation of ${\rm C_3H_5}^+$ and ${\rm C_3H_7}^+$ in gas phase.
But it is already discussed in section 2.2 that these two reactions contain high barrier and it is quite
problematic to overcome such a high obstacle around low temperature.
Moreover, due to high energy barrier of O atom ($1660$ K for ground state oxygen and supra-thermal oxygen atom), 
oxygenation reaction is not as fast as hydrogenation reaction in low temperature ($T_{ice}=10$ K) and thus ice phase PrO formation is inadequate in low temperature. 
At the low temperature ($\sim 10$ K) PrO can mainly be formed in the gas phase by the reaction between 
$\rm{O}$ atom and $\rm{C_3H_6}$,  where the major contribution of $\rm{C_3H_6}$ is coming from the ice.
As we have increased the temperature (right panel, $T_{ice}=25$ K), the situation has changed because of the increase in 
the mobility of O atom on grain which enables a significant production of 
PrO on the grain surface.
From our Dark cloud model (for $a=0.03$ and $T_{ice}=10$ K), in between our simulation time scale ($\sim 10^6$ years), 
we have a peak abundance (w.r.t. $\rm{H_2}$) of PrO to be $4.17 \times 10^{-14}$ and
$5.26 \times 10^{-14}$ respectively in gas and ice phase whereas for $a=0.03$ and $T_{ice}=25$ K it is
$1.33 \times 10^{-11}$ and $2.04 \times 10^{-7}$ respectively. In order to check the dependency of our result on the
fiducial factor, we further have considered $a=0.01$ and $T_{ice}=25$ K and found that the peak gas phase abundance 
($\sim 4.58 \times 10^{-12}$) decreased by $\sim 3$ times compared to the case with $a=0.03$ and T$_{ice}=25$ K.

Since, the abundance of $\rm{C_3H_6}$ in the gas phase is not adequate, no significant amount of 
PrO is obtained for $T_{ice}=10$ K but for $T_{ice}=25$ K, formation of $\rm{C_3H_6}$ significantly enhanced and resulting
significant increase in the production of PrO. To find out a parameter space for the formation of 
PrO in the dark cloud condition, we have varied the number density of total hydrogen ($n_H$) 
in between $10^4-10^7$ cm$^{-3}$ and ice temperature between $10-25$ K and kept the gas temperature at $30$ K. 
The resulting plot is shown in Fig. 3. It is clear from the plot that the observed abundance 
{($\rm{\sim\times10^{-11}}$)} of PrO can be obtained in between  $n_H=10^4-10^5$ cm$^{-3}$ 
and ice temperature range $20-25$ K.

Since, we have considered a small activation barrier of about $40$ K for the ice phase reaction between
$\rm{C_3H_6}$ and O($^3$P), consideration of supra-thermal oxygen has not influenced our results. It would be kept in
mind that due to the unavailability of the binding energy values for O($^1$D), we have considered the same
binding energy as it was for O($^3$P). However, it is expected that due to the generation process of the supra-thermal oxygen,
it would carry some extra energy which would enable it to move much faster than O($^3$P). Thus, it is expected that
the binding energy of
O($^1$D) would be lower than O($^3$P).
We have tested the abundance of PrO again by considering a lower binding energy value (half of the normal oxygen) for O($^1$D). However, again for this case, we have not found any significant difference. The reason behind is that in one hand lower
binding energy provides much faster diffusion rate and enhances the chance of recombination, but
on another hand, it gives much quicker sublimation rate which decreases the residence
time of O($^1$D) on the grain surface and thus reduces the chance of recombination. 
Since in our network, PrO mainly formed by the reaction between $\rm{C_3H_6}$ and O($^3$P), lowering the binding energy 
of O($^1$D) have not changed our PrO abundance.
A higher dose of cosmic rays would be useful to generate more supra-thermal oxygen which could be helpful in the PrO formation, but it would also dissociate PrO as well.   
We think that the cosmic ray ionization rate of about $1.3 \times 10^{-17}$ s$^{-1}$ which is used here,
is a standard value for the region where the three transitions of PrO had been observed.

\begin{figure}
{\centering
\includegraphics[width=9cm]{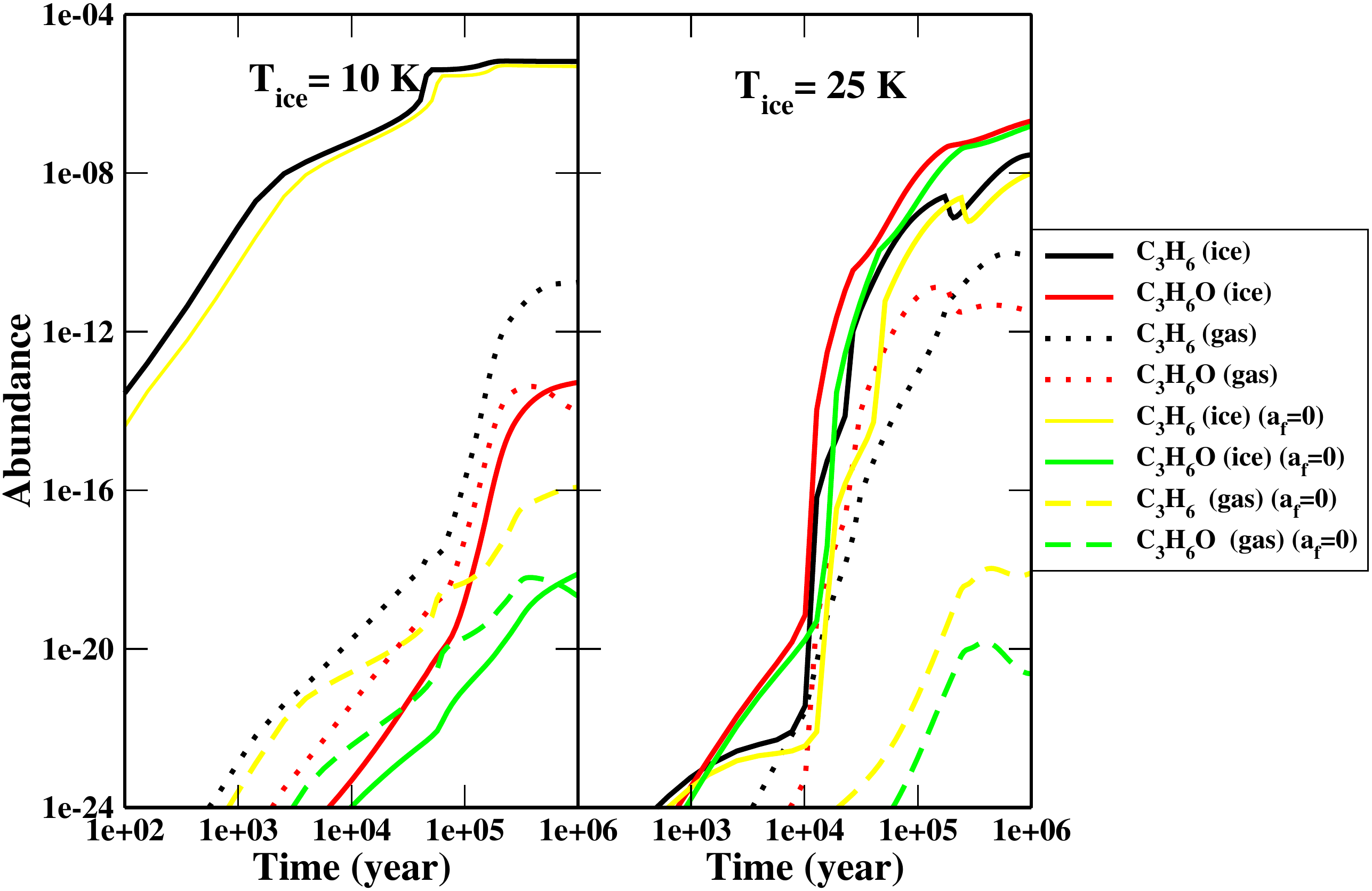}
\caption{Chemical evolution of PrO and $\rm{C_3H_6}$ for a dark cloud model. Solid line 
represents ice phase abundance and dotted line represents the gas phase abundance.}}
\end{figure}
\vskip 0.5cm
\begin{figure}
{\centering
\includegraphics[width=6cm,angle=270]{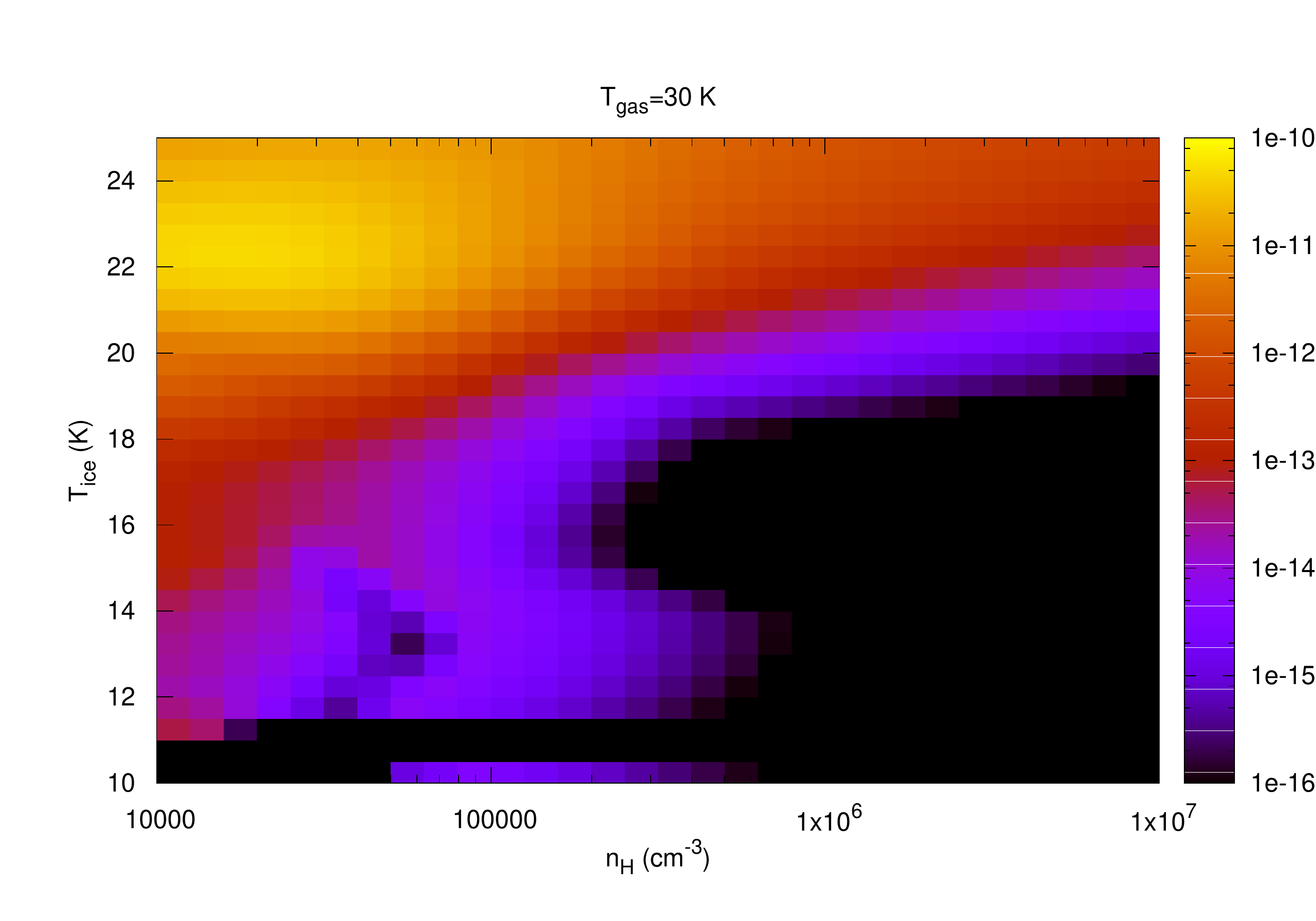}
\caption{Parameter space of the abundance of PrO in dark cloud conditions. Color bar represents the fraction abundance of PrO}}
\end{figure}

\begin{table}
{\scriptsize
\centering
\caption{Vibrational analysis of propylene oxide using B3LYP/6-311++G(d,p)}
\begin{tabular}{|p{0.25in} p{0.25in}|p{0.75 in}|p{1.158in}|p{0.25in}|p{0.25in}|c}
\hline
{\bf Wavenumber} & &{\bf Integral Absorbance} & {\bf Band assignment} & {\bf Exp. value }\\
\cline{1-2}
$\nu$(cm$^{-1}$)& $\lambda$($\mu$m) &{\bf Coefficient(cm molecule$^{-1}$)} &&\citep{huds17}\\
\hline
218.85&45.84&1.31$\times10^{-19}$&CH$_3$ torsion&\\
367.62&27.20&1.23$\times10^{-18}$&CCC bending&\\
411.41&24.30&1.19$\times10^{-18}$&CCO bending &\\
752.24&13.29&2.47 (1.43)$^{a}$$\times10^{-18}$&ring deformation&742.0\\
826.19&12.10&15.6 (9.53)$^{a}$$\times10^{-18}$&ring deformation&825.9\\
905.22&11.04&8.69 (4.8)$^{a}$$\times10^{-19}$&CH$_2$ rocking&896.7\\
961.52&10.40&4.23 (2.35)$^{a}$$\times10^{-18}$&CC stretching or CH$_3$ rocking&948.3\\
1038.38&9.63&3.08 (2.46)$^{a}$$\times10^{-18}$&CH$_3$ wagging&1027.3\\
1119.31&8.93&2.01$\times10^{-18}$&CH$_2$ rocking&\\
1149.97&8.69&8.84$\times10^{-19}$&CH$_2$ wagging&\\
1163.75&8.59&8.85$\times10^{-19}$&CH$_3$ bending&\\
1185.69&8.43&2.54$\times10^{-19}$&CH bending&\\
1291.89&7.74&1.52(1.47)$^{a}$$\times10^{-18}$&CH bending, ring deformation &1266\\
1401.63&7.13&7.90(2.57)$^{a}$$\times10^{-19}$&CH$_3$ deformation&1408\\
1434.36&6.97&5.23$\times10^{-18}$&CC stretching&\\
1473.10&6.78&1.70$\times10^{-18}$&CH$_3$ deformation&\\
1486.40&6.72&1.78$\times10^{-18}$&CH$_3$ deformation&\\
1520.59&6.57&1.98$\times10^{-18}$&CH$_2$ scissoring&\\
3027.69&3.30&3.98$\times10^{-18}$&CH$_3$ symmetric stretching&\\
3087.79&3.24&5.92$\times10^{-18}$&CH$_2$ asymmetric stretching&\\
3089.63&3.23&4.84$\times10^{-18}$&CH$_2$ symmetric stretching&\\
3098.32&3.22&1.12$\times10^{-18}$&CH stretching&\\
3116.85&3.20&1.02$\times10^{-17}$&CH stretching&\\
3179.24&3.14&5.70$\times10^{-18}$&CH$_2$ asymmetric stretching&\\
\hline
\end{tabular}
$^a$ \cite{huds17}.}
\end{table}

\begin{table}
{\scriptsize
\centering
\caption{Vibrational analysis of protonated propylene oxide ($\rm{C_3H_6OH^{+}}$) using B3LYP/6-311++G(d,p)}	
\begin{tabular}{|c c|c|c|c|c|c}
\hline
{\bf Wavenumber} & &{\bf Integral Absorbance} & {\bf Band assignment}\\
\cline{1-2}
$\nu$(cm$^{-1}$)& $\lambda$($\mu$m) &{\bf Coefficient(cm molecule$^{-1}$)} & \\
\hline
207.33&48.23&8.78$\times10^{-20}$&CH$_3$ torsion\\
325.32&30.73&4.77$\times10^{-18}$&CCO bending\\
395.82&25.26&2.25$\times10^{-18}$&CCC bending \\
494.41&20.22&9.16$\times10^{-18}$&ring deformation\\
748.95&13.35&2.56$\times10^{-17}$&CO stretching \\
845.94&11.82&1.06$\times10^{-17}$&OH torsion \\
894.73&11.17&1.03$\times10^{-17}$&CC stretching \\
936.21&10.68&1.48$\times10^{-18}$&CH$_3$ wagging, CH$_2$ torsion\\
982.36&10.17&5.08$\times10^{-18}$&CH$_2$ rocking, OH bending\\
1034.35&9.66&6.07$\times10^{-18}$&CH$_3$ wagging\\
1134.12&8.81&1.72$\times10^{-18}$&CH, CH$_3$ bending \\
1212.99&8.24&5.40$\times10^{-19}$&CH$_2$ wagging\\
1225.73&8.15&6.52$\times10^{-18}$&CH, OH bending\\
1231.93&8.11&1.33$\times10^{-18}$& CH$_3$ bending\\
1303.36&7.62&2.82$\times10^{-18}$&CC stretching\\
1404.63&7.11&1.43$\times10^{-18}$&CH$_3$ deformation\\
1432.85&6.97&4.37$\times10^{-18}$&CH$_3$ deformation\\
1463.20&6.83&3.78$\times10^{-18}$&CH$_3$ deformation\\
1492.42&3.30&6.70$\times10^{-18}$&CH$_3$ deformation\\
1512.84&3.24&6.61$\times10^{-18}$&CH$_2$ scissoring\\
3045.28&3.28&6.32$\times10^{-20}$&CH$_3$ symmetric stretching\\
3113.53&3.21&5.32$\times10^{-19}$&CH$_3$ asymmetric stretching\\
3146.53&3.17&7.45$\times10^{-19}$&CH$_2$ asymmetric stretching\\
3165.61&3.15&5.55$\times10^{-20}$&CH$_2$ symmetric stretching\\
3201.52&3.12&1.5$\times10^{-19}$&CH stretching\\
3274.58&3.05&5.14$\times10^{-19}$&CH$_2$ asymmetric stretching\\
3675.89&2.74&5.45$\times10^{-17}$&OH stretching\\
\hline
\end{tabular}}
\end{table}

\subsection{Hot core model}
\subsubsection{Physical condition}
\cite{cunn07} attempted to observe PrO and glycine in Sgr B2(LMH) and Orion-KL by using MOPRA telescope. 
However, they did not detect either species but were able to put an upper limit on the abundances of these two molecules in these sources.
They proposed an upper limit of about $6.7 \times 10^{14}$ cm$^{-2}$ for the column density of PrO 
in Sgr B2 (LMH).  Here, to estimate the abundance of PrO in the hot core region, 
we have considered a two-phase model as used by \citet{garr06}. The first phase is assumed to be isothermal at $10$ K  or $25$ K and 
lasts for $10^6$ years. For this phase, 
we have considered $A_V=30$ and $\zeta=1.3 \times 10^{-17}$ s$^{-1}$.
The subsequent phase is assumed to be a warm-up period where the temperature can gradually increase { up to} 
$200$ K in $10^5$ years. 
So our total simulation time is restricted up to a total $1.1 \times 10^6$ years. The number density of the 
total hydrogen is assumed to be constant ($n_H=10^4-10^7$ cm$^{-3}$)
in both the phases of our simulation. 

\subsubsection{Results \& Discussions}
In Fig. 4, we have shown the chemical evolution of PrO for various density clouds 
($n_H=10^4 - 10^7$ cm$^{-3}$). 
The last panel shows the abundance variation of peak gas phase abundance and final abundance (i.e., at the end of the simulation 
the time scale $\sim 1.1 \times 10^6$ year) of PrO with the number density variation.
In case of the $T_{ice}=10$ K, as we have decreased the density, 
gas phase abundance of PrO increased slightly and have a maximum ($\sim 1.5 \times 10^{-7}$ w.r.t. ${\rm H_2}$)
around $n_H=1 \times 10^4$ cm$^{-3}$. In the high density region ($\sim 10^7$ cm$^{-3}$), we have obtained 
the peak abundance $\sim 1.3 \times 10^{-9}$. The final abundance for this case, varies in between 
$1.4\times 10^{-10}-8.9 \times 10^{-8}$. In case of $T_{ice}=25$ K, the peak and final abundance roughly remain 
invariant (peak abundance varies in between $1.03 \times 10^{-7}-2.4 \times 10^{-7}$ and final abundance varies in between 
$1.7 \times 10^{-8}-3.5 \times 10^{-8}$). 

\begin{figure}
{\centering
\hskip -1.2cm
\includegraphics[height=5cm,width=10cm]{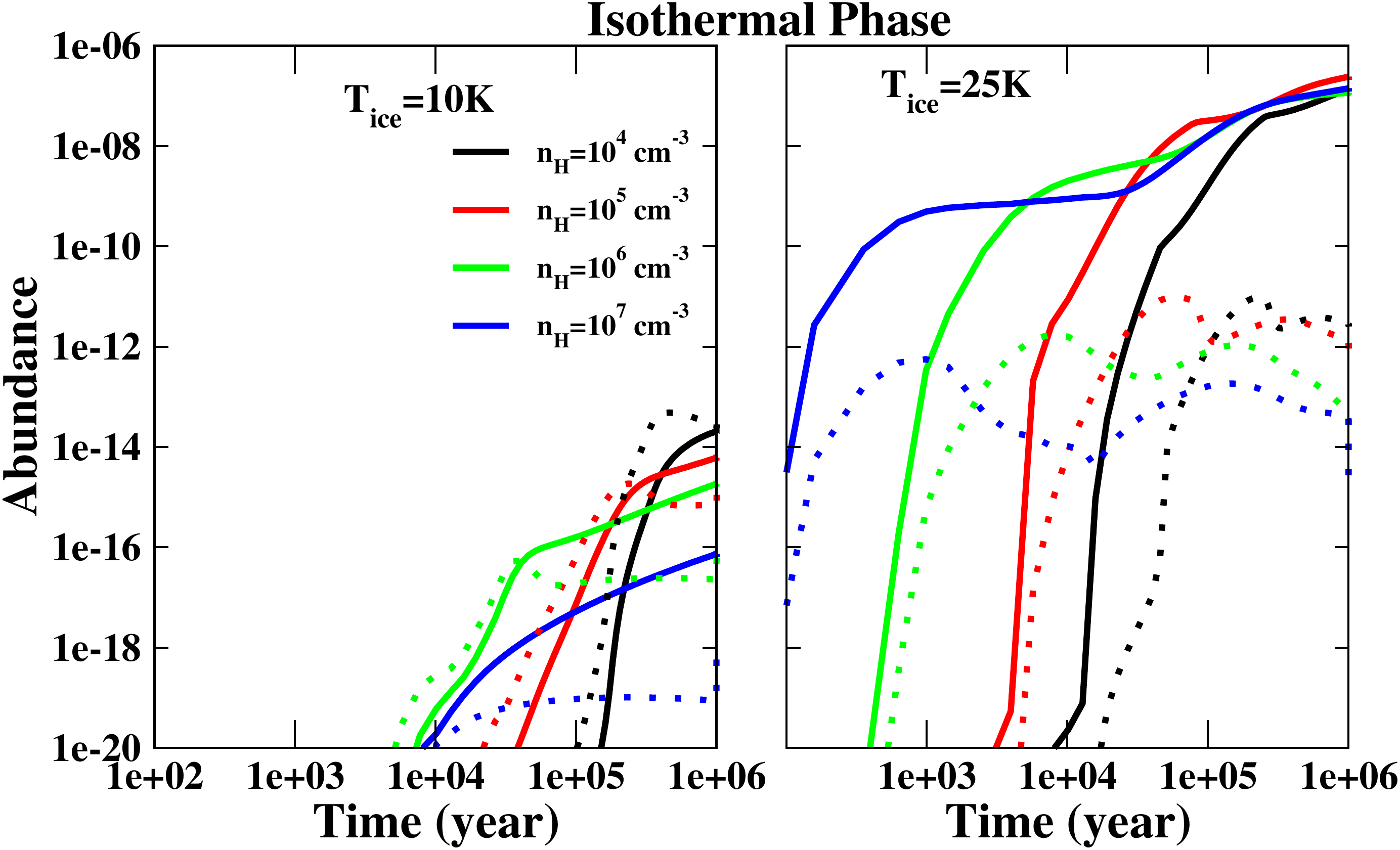}
\vskip 0.5cm
\hskip -1.2cm
\includegraphics[height=9cm,width=10cm]{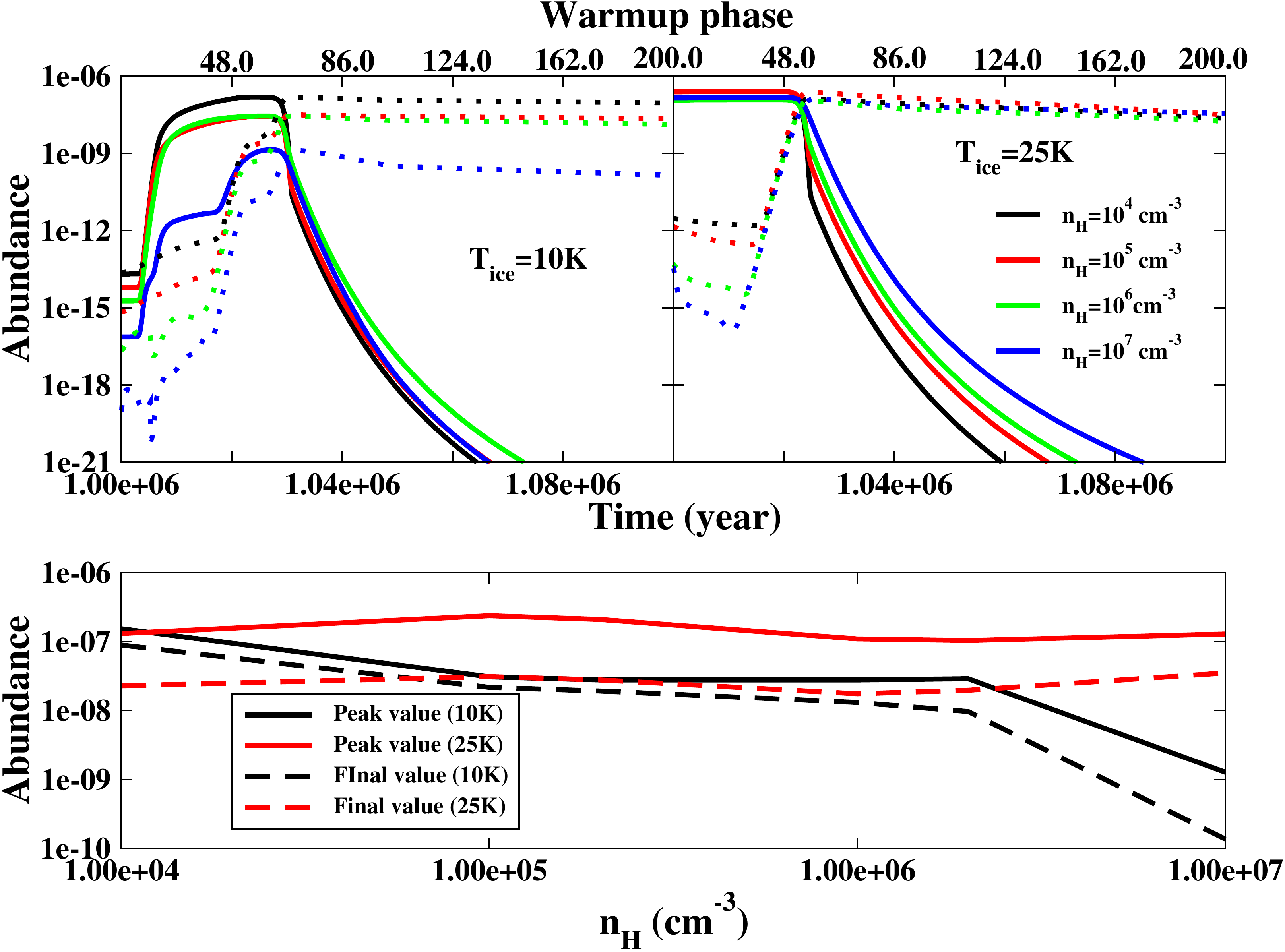}
\caption{Chemical evolution of PrO in a Two phase model by considering various density cloud. 
The upper two panels are for the isothermal phase, middle two panels are for the warm-up, and the lower panel is for the gas phase peak 
abundance of PrO and final abundance of PrO at the warm-up phase for various density cloud. Different colors are used to denote different 
density cloud.
In the upper and middle panels, the solid line represents the abundance in the gas phase
whereas the dotted line represents the abundance in ice phase.}
}
\end{figure}

\subsection{Comparison with the previous modeling results}
There are some basic differences between the model described above and used by the other authors to
explain the observed abundance of PrO.
Recently, \cite{berg18} used a combined experimental and theoretical study to explain the abundance of PrO in the ISM.
They explained the formation of PrO by non-equilibrium reactions initiated by the effects of secondary
electrons. More specifically, they used supra-thermal ($^{1}$D) oxygen insertion reaction with propylene to explain the
PrO formation. They did not consider the formation of PrO by the addition of ground state oxygen ($^{3}$P).
In our case, we have considered the formation of PrO by both types of oxygen atoms. \cite{ward11} found a low activation barrier ($\sim 40$K) for the ice phase reaction between
$\rm{C_3H_6}$ and O($^{3}$P). Such a low activation barrier could be overcome at the low temperature. As a results, we
have found an adequate production of PrO even in absence of the supra-thermal oxygen insertion reaction.

\section{Vibrational Spectroscopy}
\subsection{Methodology}
To compute the vibrational transitions of PrO and its protonated form, we have 
used Gaussian 09 program. We have used the 
Density Functional Theorem (DFT) with 6-311G basis set including diffuse and polarization functions for this
calculations. Initially, we have 
verified our results with various methods and basis sets and compared it with the experimentally obtained results. 
We have found that 
B3LYP/6-311++G(d, p) method is best suited to reproduce the experimentally obtained transitions of PrO. Moreover, we also have considered 
IEPCM model with the method mentioned above to compute the vibrational transitions of PrO in the ice phase. All optimized geometry was 
verified by harmonic frequency analysis (having no negative frequency). We have performed our calculations by placing the solute (species) 
in the cavity within the solvent (water molecule) reaction field. Here, we have used the SCRF method by following the earlier study of 
\cite{woon02}.

\subsection{Results and discussions}
In Table 5, we have summarized calculated and experimental vibrational frequencies of PrO along with the integral 
absorption coefficients and band assignments. Similarly, Table 6 presents the determined vibrational frequencies 
along with the integral absorption coefficient of protonated PrO.
Our calculated values of absorption coefficients and vibrational frequencies are in excellent agreement with the experimentally obtained results 
\citep{huds17}. \cite{puzz14} studied the spectroscopic details of protonated oxirane ($\rm{C_2H_5O^{+}}$). Similarly, protonation of PrO can take 
place. PrO could be protonated via ion-neutral reactions where the proton is transferred to the neutral. Protonated PrO may be formed in the gas 
phase by the reaction of PrO with the ions like $\rm{H_3O^{+}}$, HCO$^{+}$ and $\rm{H_3^{+}}$. We have found a proton affinity of PrO about 
$803.169$ kJ/mol by doing quantum chemical calculations (using the HF/6-31G(d) method) which is in an excellent agreement with \cite{hunt98} ($803.3$ kJ/mol). For another 
protonated form of PrO, proton affinity value is found to be $801.827$ kJ/mol. Thus, protonated PrO shown in Fig. 1c possessing
a slightly higher energy (about $175$ K) than protonated PrO shown in Fig. 1b. 
Since the experimental data for protonated PrO is yet to be published, based on the accuracy obtained for the vibrational 
transitions of PrO, same method and basis set have been applied for protonated PrO to compute the vibrational transitions and integral absorption 
coefficients.
Figure 5ab depicts the absorption spectra of PrO and its protonated form.
Figure 5a covers $1600-200$ cm$^{-1}$ region ( some parts of the mid IR: $2000-400$ cm$^{-1}$ and some parts of the 
far IR: $400-50$ cm$^{-1}$) 
and Fig. 5b covers $3800-3000$ cm$^{-1}$ region (i.e., some parts of the near IR: $12800-2000$ cm$^{-1}$).

\subsection{Comparison with the experimentally obtained results}
In the top and bottom panel of Fig. 6a comparison between our computed and experimentally obtained vibrational spectra is shown 
for ($1600-800$) cm$^{-1}$ and ($3200-2700$) cm$^{-1}$ frequency window respectively. 
In these two panels, we also have shown how the clustering of PrO can affect the computed vibrational spectra of PrO. 
The blue line in Fig. 6 represents the experimental
data extracted from \cite{huds17} whereas the black, red, and green line represent the calculated spectra with PrO monomer, PrO dimer, and PrO 
tetramer respectively. Fig. 6a depicts that in the mid-IR region,
we have a good agreement between our theoretical and experimental results of \cite{huds17}. 
We have noticed that most of the transitions within this mid-IR region are
within an error bar of about $10$ cm$^{-1}$. One transition at $1291.89$ cm$^{-1}$ is found to be 
shifted maximum by $25$ cm$^{-1}$. The intensity of each mode of vibration and the area under the curve of various
regime changes with the cluster size of PrO. We have noticed that if a scaling factor of $0.9728$ is used, we have 
an excellent agreement around the stretching mode and thus in Fig. 6b, we have scaled our computed wavenumber accordingly.

\begin{figure}
\includegraphics[width=8cm]{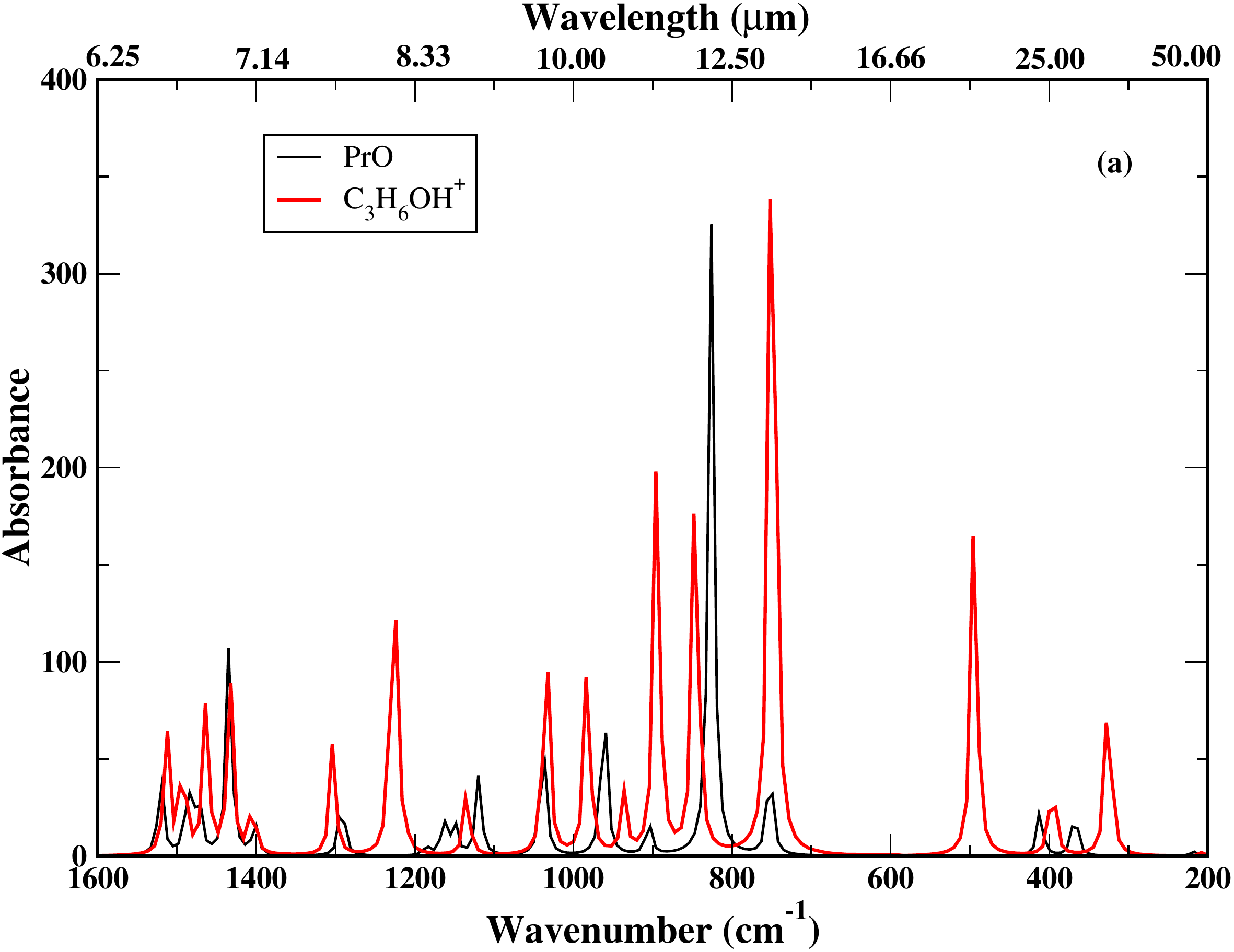}
\vskip 0.5cm
\includegraphics[width=8cm]{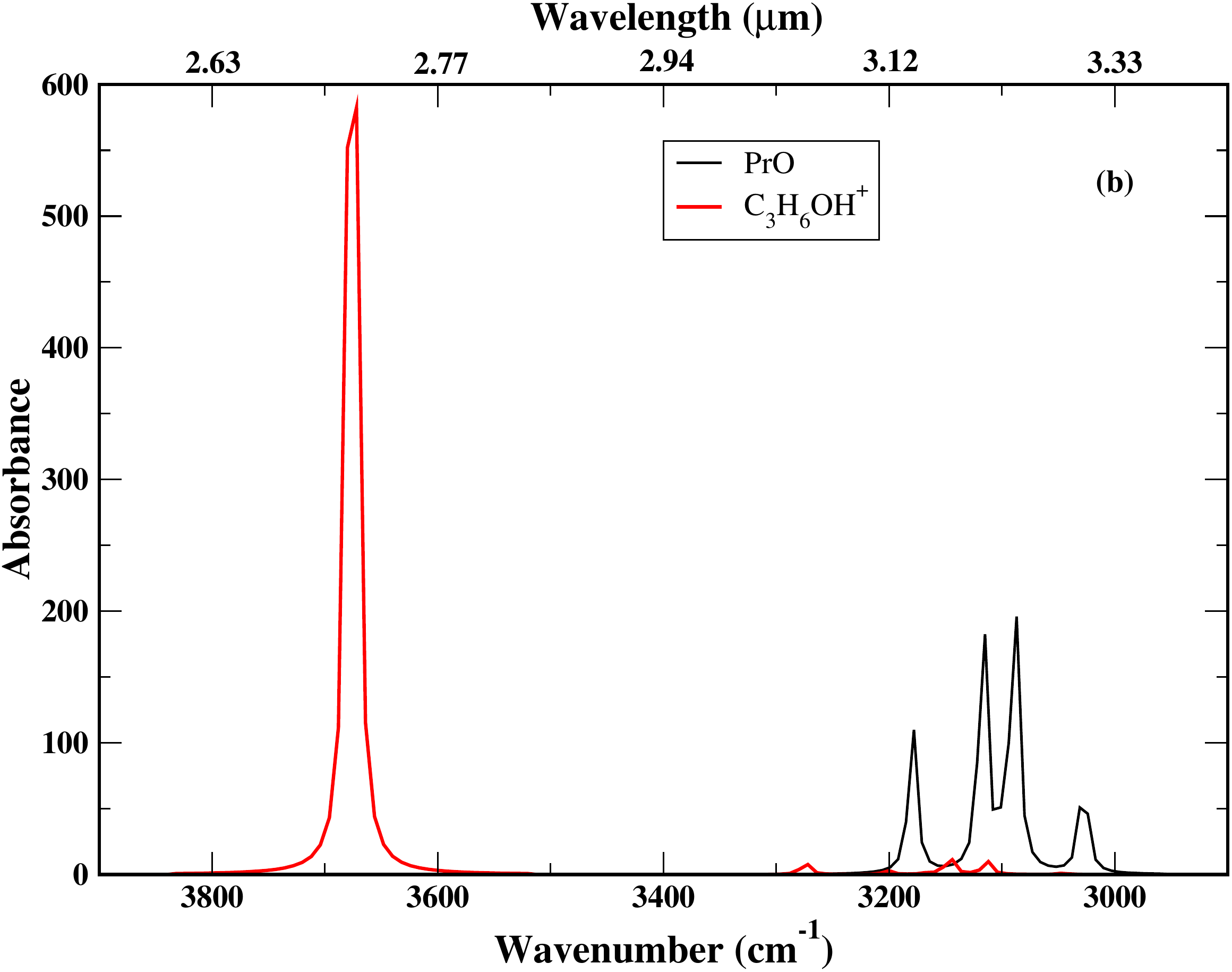}
\caption{Comparison between our calculated IR spectra of Propylene Oxide and protonated propylene oxide.}
\end{figure} 

\begin{figure}
\includegraphics[width=8cm]{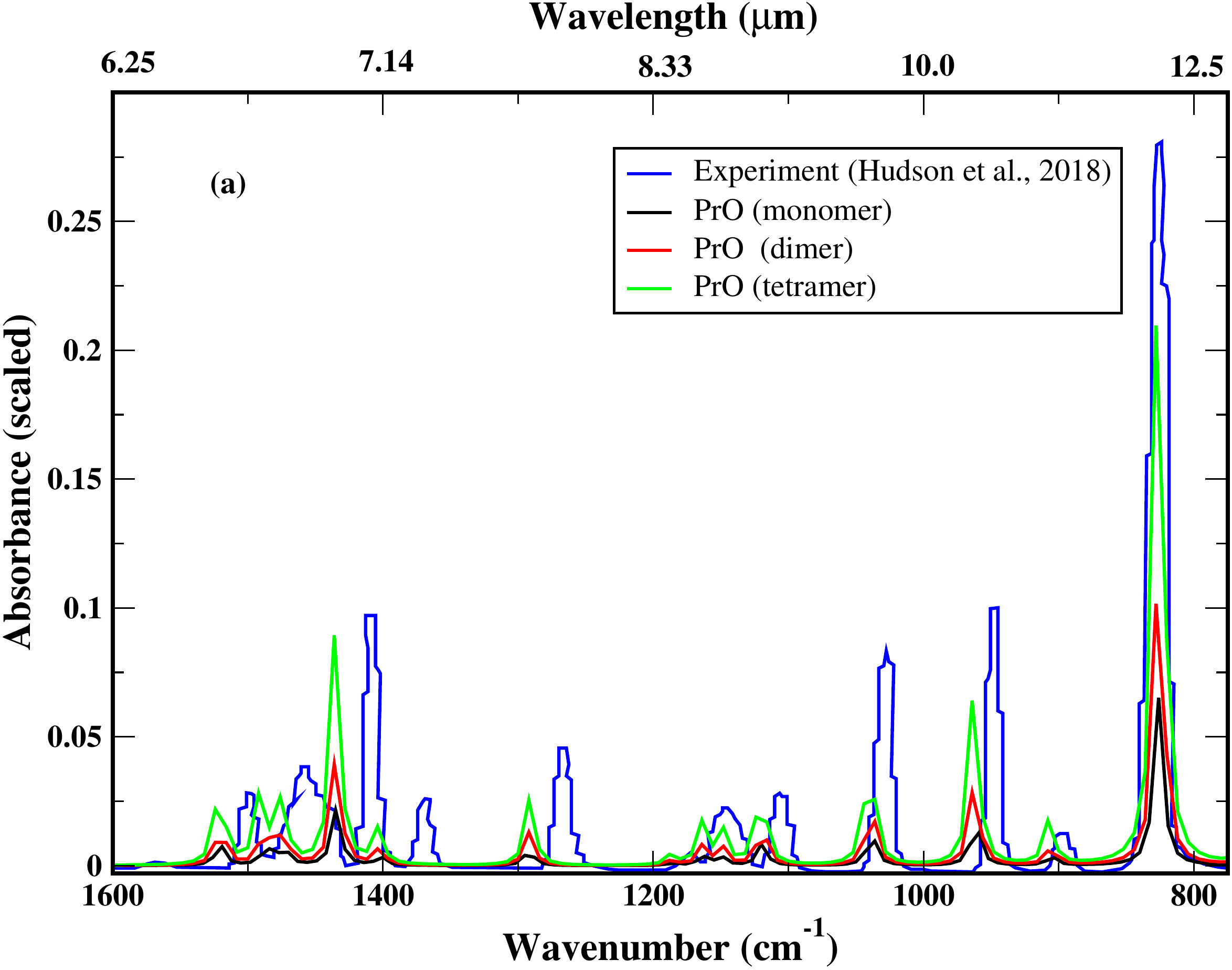}
\vskip 0.5cm
\includegraphics[width=8cm]{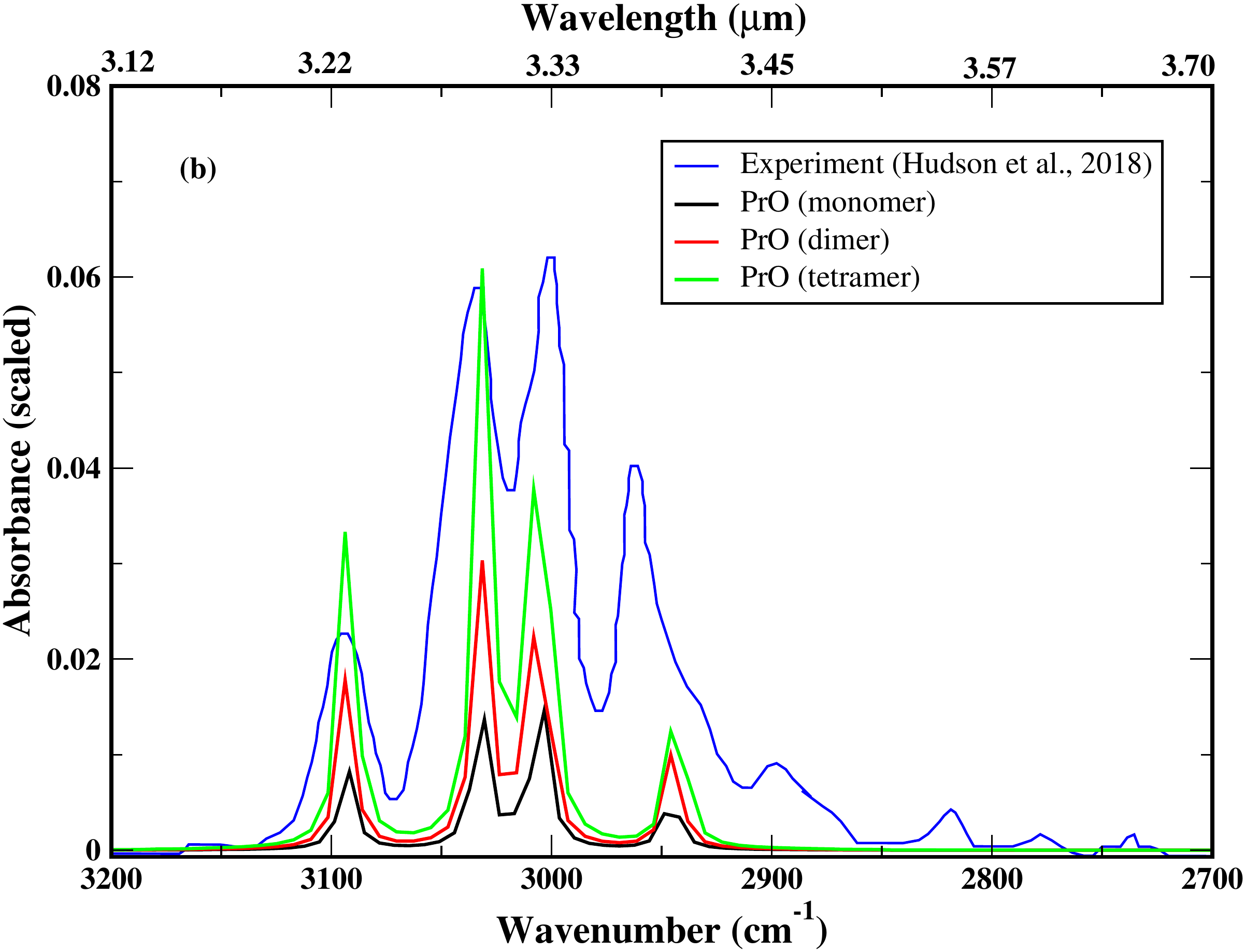}
\caption{Comparison between our calculated spectra with that obtained in the experiment \citep{huds17}.
{ Results of \citep{huds17} has been digitally extracted by using
https://apps.automeris.io/wpd. In the upper panel, we have shown the mid-IR region 
and in the bottom panel, we have shown the near-IR region. For the sake of better visualization,
we have scaled down the Y axis of our calculated values in (a) and (b). Our calculated
wavenumber in X-axis of (b) is scaled by $0.9728$ to have better agreement.}
}
\end{figure}

\begin{figure}
\centering
\includegraphics[height=8cm, width=9cm]{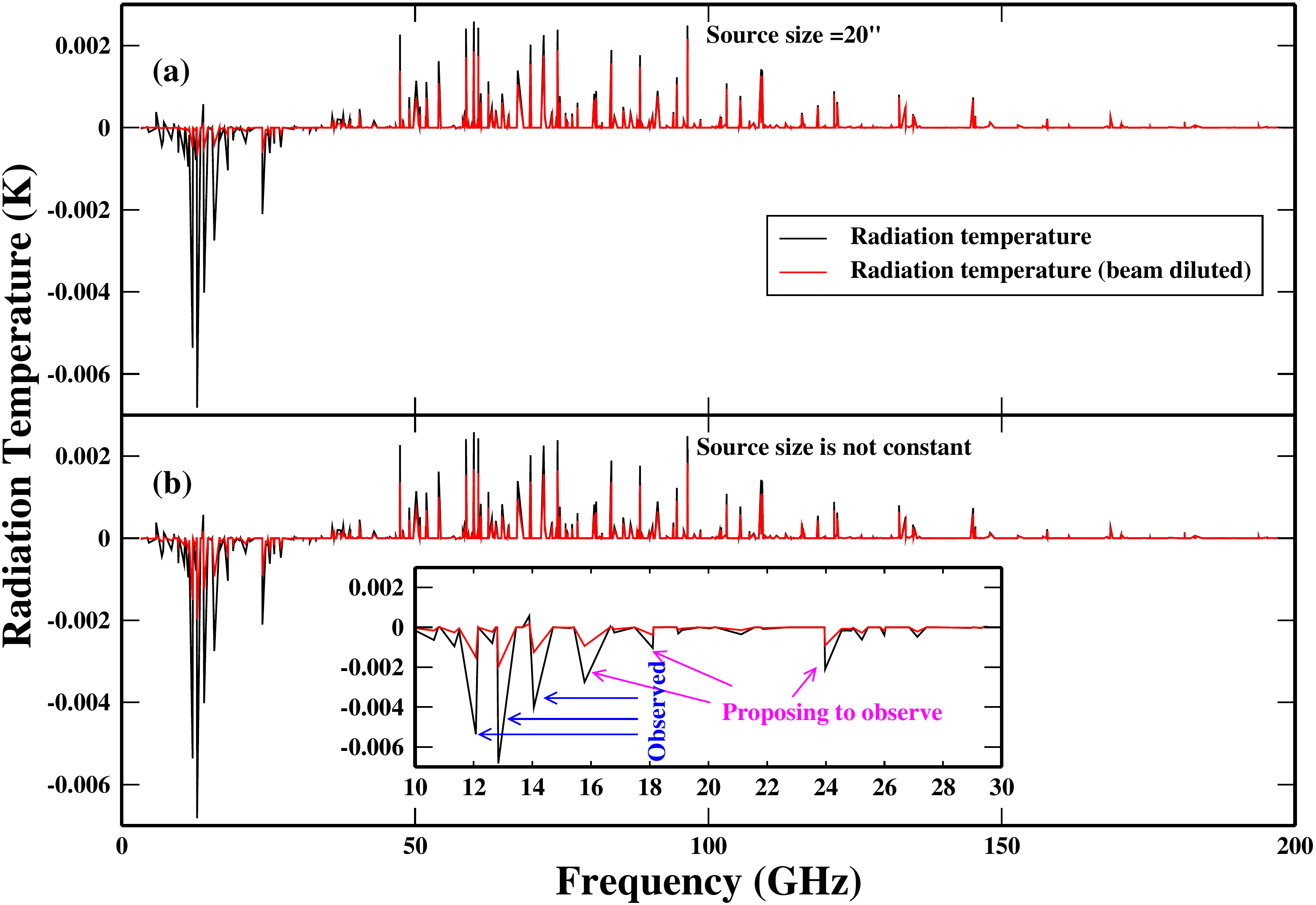}
\caption{Line parameters of PrO for non-LTE condition by considering (a) constant source size ($20"$) and (b) varied source size \citep{holl07}.
Observed and other possible transitions are highlighted inside a separate box in Fig. 7.}
\end{figure}

\section{Radiative transfer modeling}
\subsection{Methodology}
In Astrochemistry, a major aim of radiative transfer modeling is to extract the
molecular abundances from the line spectra at infrared and sub(millimeter) wavelengths.
Three different levels of radiative transfer model are in use which ranges from
the basic Local Thermodynamic Equilibrium (LTE) models to complex non-local
models in 2D and even more complex models in 3D \citep{vand11}.

A simple single excitation temperature model, i.e., LTE modeling has been carried out
to predict the most probable transition of PrO.
LTE modeling is considered to be a good starting point for predicting the line parameters around
the high-density region. Moreover, it does not require additional collisional
information. We have used CASSIS program which was developed by IRAP-UPS/CNRS 
\url{(http://cassis.irap.omp.eu)} for the LTE modeling. 
Recently, \cite{mcgu16} observed $3$ transitions of PrO in absorption. 
\cite{cunn07} used the MOPRA telescope at the $3$ mm band for the
identification of various transitions of PrO in Sgr B2(LMH) and Orion-KL.
However, they were unable to detect any PrO transitions in these two targeted regions but
they had provided an upper limit for the abundance of PrO around these sources.
We have carried out the LTE calculations for two different cases. First one 
is for the sources where we are expecting the transitions of PrO in the extended molecular shell
around the embedded, massive protostellar object as observed by \cite{mcgu16} and for the second one,
we have considered the input parameters in such a way that we may observe
some strong transitions of PrO in the hot core regions as attempted by \cite{cunn07}.\\

In the LTE model, level populations are controlled by Boltzmann distribution.
As Interstellar chemical process far
from thermodynamic equilibrium, it is essential to consider the non-LTE method where the balance between the excitation
and de-excitation of molecular energy levels are explicitly solved. Collisional and radiative both
the processes contribute to this balance. Most importantly, for the non-LTE method, the collisional
data file is required which is not available for most of the species.
Collisional data file for PrO was not available in any database.
To have an educated estimation of the line parameters in non-LTE, we have considered the
collisional rate parameters of $\rm{e-CH_3OH}$ with H$_2$ \citep{rabl10} for preparing the collisional data file of
PrO.
Since PrO is a slightly asymmetric rotor molecule, it would have been better if we have used the collisional
parameters of $\rm{H_2CO}$ here. But due to the unavailability of the collisional rates for the
sufficient number of energy levels for $\rm{H_2CO}$, tempted us to use $\rm{e-CH_3OH}$ as an approximation for which we have the collisional parameters available in LAMDA database \citep{scho05} for sufficient number of energy levels.
RADEX program \citep{vand07} have been used for this non-LTE computation.

\begin{table}
\centering{ \scriptsize
\caption{Input parameters used for the LTE modeling}
\begin{tabular}{c|c}
\hline
\multicolumn{2}{c}{\bf Parameters used for GBT (100 m)$^{a}$}\\
\hline
Column density of H$_2$&10$^{24}$ cm$^{-2}$\\
Column density of propylene oxide& $1.0 \times 10^{13}$ cm$^{-2}$\\
Excitation temperature $(T_{ex})$ & 5 K\\
Background temperature $(T_{C})$ & 2.73 K (default for LTE)\\
FWHM &$15$ km/s\\
Source size ($\theta_s$) &$20''$\\
Beam size ($\theta_b$)& $\frac{744}{\nu}$\\ 
Beam dilution & B=$\frac{\theta_s^{2}}{(\theta_s^{2}+\theta_b^{2})}$\\
\hline
\multicolumn{2}{c}{\bf Parameters used for ALMA (400 m)$^{b}$}\\
\hline
Column density of H$_2$&10$^{24}$ cm$^{-2}$\\
Column density of propylene oxide& $1.74 \times 10^{16}$ cm$^{-2}$\\
Excitation temperature $(T_{ex})$ & 150 K\\
Background temperature $(T_{C})$ & 2.73 K (default for LTE)\\
FWHM &$5$ km/s\\
Source size &$3.0''$\\
Beam Dilution&$1$\\
\hline
\end{tabular}}

{\scriptsize
$^{a}$ GBT model parameters are based on \cite{mcgu16}.\\
$^{b}$ALMA parameters are based on the model of this work.}
\end{table}

\begin{table}
\centering
\scriptsize{
\caption{LTE line parameters of the various transitions of PrO using GBT.}
\begin{tabular}{|c|c|c|}
\hline
Frequency (GHz)& $J_{Ka'kc'}-J_{ka''kc''}$&Intensity (mK)\\
\hline\hline
12.07243$^o$& $1_{10}-1_{01}$ & $9.47\times 10^{-6}$\\
12.83734$^o$& $2_{11}-2_{02}$ & $1.83\times 10^{-5}$\\
14.04776$^o$& $3_{12}-3_{03}$ & $2.82\times 10^{-5}$\\
\hline
\end{tabular}\\
\hskip 5cm {\scriptsize $^o$\cite{mcgu16}}}\\
\end{table}

\subsection{Results and discussions}
\subsubsection{LTE model}
For the first case, we have considered the GBT $100$m telescopic parameter for the modeling and adopted
parameters are shown in Table 7. For this case, we have assumed that the source size ($\theta_s$)
$20''$ and beam size ($\theta_b$) may vary depending upon the frequency by $\frac{744}{\nu}$ \citep{holl07}, where 
$\nu$ is the frequency in GHz. Beam dilution factor is computed by $B=\frac{{\theta_s}^2}{({\theta_s}^2+{\theta_b}^2)}$ and 
applied on the obtained intensity. The recent observation
by \cite{mcgu16} observed three transitions of PrO in cold molecular shell in front of the bright continuum
sources/hot cores within Sgr B2. In Table 8, we have pointed out the obtained LTE parameters for these three transitions only.
We have noticed that the obtained intensities are very weak and are even below the RMS noise level of 1 mK.
The reason behind this result is that in the CASSIS module of LTE, we have used a constant background continuum temperature
of $2.73$ K. However, a variable background continuum temperature was required to observe these transitions in absorption.
We will get back to the discussion about the absorption feature of these three transitions in the later section of this paper again.

For the second case, we have modeled the LTE transitions of PrO for the hot core region.
Here, we have used our computed gas phase PrO abundance $\sim\rm{1.74\times10^{-8}}$
(final abundance with $n_H=10^6$ cm$^{-3}$ or $n_{H_2}= 5 \times 10^5$ cm$^{-3}$ with initial dust temperature $25$ K). We have used ALMA 400 m telescopic parameter 
for the modeling of this feature.
Here, we have considered source size ($1.5''$), beam size varies as $\frac{186}{\nu}$ (for ALMA 400m) and the beam dilution effect 
applied following the same formula ($B=\frac{{\theta_s}^2}{({\theta_s}^2+{\theta_b}^2)}$) mentioned above.
To verify the potential observability of these PrO transitions, we have checked 
the blending of these transitions with any other interstellar molecular transitions.
Interestingly, we have identified a few intense transitions of PrO in Band 3 ($84-116$ GHz) 
and Band 4 ($125-163$ GHz) which are not blended.
In Table 9, we have pointed out all the potentially observable transitions of PrO in the hot core region.
Based on our simple
LTE model, we have isolated some strong transitions of PrO.
We further have used ALMA simulator to check the integration time required to observe these transitions by ALMA.
We have found that for ALMA Band 4 (for e.g 130 GHz), with $20$ mK RMS noise, $0.4$ km/s spectral resolution and
$43$ antennas along with the dual polarization, $1$ hr on-source integration time is required
for 5$\sigma$ detection of PrO in Sgr B2 (N) with an angular resolution of $1.5''$. For the similar configuration,
$3$ hrs on-source integration time is required for Band 3 (e.g 100 GHz) observation with ALMA.\\

\subsubsection{non-LTE}
In Fig. 7ab, we have shown our computed radiation temperature for the various transitions of PrO within
a wide range of frequency ($1-200$ GHz). 
Here, our target was to find out the line parameters which are relevant for the observations performed by
\cite{mcgu16}. Thus, we have considered, $T_{ex}=5$ K, column density of PrO =$1 \times 10^{13}$ cm$^{-2}$, FWHM =$15$ Km/s 
(for the observed three transitions, \cite{mcgu16} obtained a FWHM of $11.6$, $15.8$ and $19.6$ respectively 
and here, we have considered the average of these three FWHMs) and
$n_{H2}=10^5$ cm$^{-3}$. For this non-LTE model, we have used GBT $100$m telescopic parameters. Background continuum
temperature measurement was already carried out by \cite{holl07}. 
They showed the observed continuum antenna temperatures by the GBT spectrometer toward SgrB2 (N-LMH). 
Here, we have considered this variation of background temperature by digitally extracting Fig. 1 
of \cite{holl07} by using \url{https://apps.automeris.io/wpd}. 
According to the Fig. 1 of \cite{holl07} the changes in continuum temperature beyond $40$GHz is very small. 
Extracting the values from \cite{holl07}, we have the continuum temperature $\sim3.35$K at $49$GHz. 
Beyond this, they did not provide any data for the continuum temperature, so we have considered a 
fixed continuum temperature of about $2.73$K beyond $100$GHz. In between the $49$GHz to $100$GHz 
we have interpolated the continuum temperature from $3.35-2.73$K. 
We have considered the variation of beam size, by considering the
relation $\sim \frac{744}{\nu}$ \citep{holl07}. In Fig. 7a, we have considered that source size is constant
at $20''$ and in Fig. 7b, we have considered the source size variation by using $\theta_s=\frac{143}{\nu^{0.52}}$ \citep{holl07}.
Radiation temperatures of the various transitions of PrO have been presented in Fig. 7ab 
by considering the beam dilution effect and avoiding this effect. 
Most interestingly, with the given conditions, we have obtained
all the three transitions ($12.07$, $12.8$ and $14.04$ GHz) in absorptions. Additionally, we have identified three
more transitions at $15.78$ GHz, $18.1$ GHz and $23.98$ GHz
which might potentially be observed around the same region, where the other three
transitions were observed.

\begin{figure}
\centering
\includegraphics[height=9.5cm,width=7cm,angle=270]{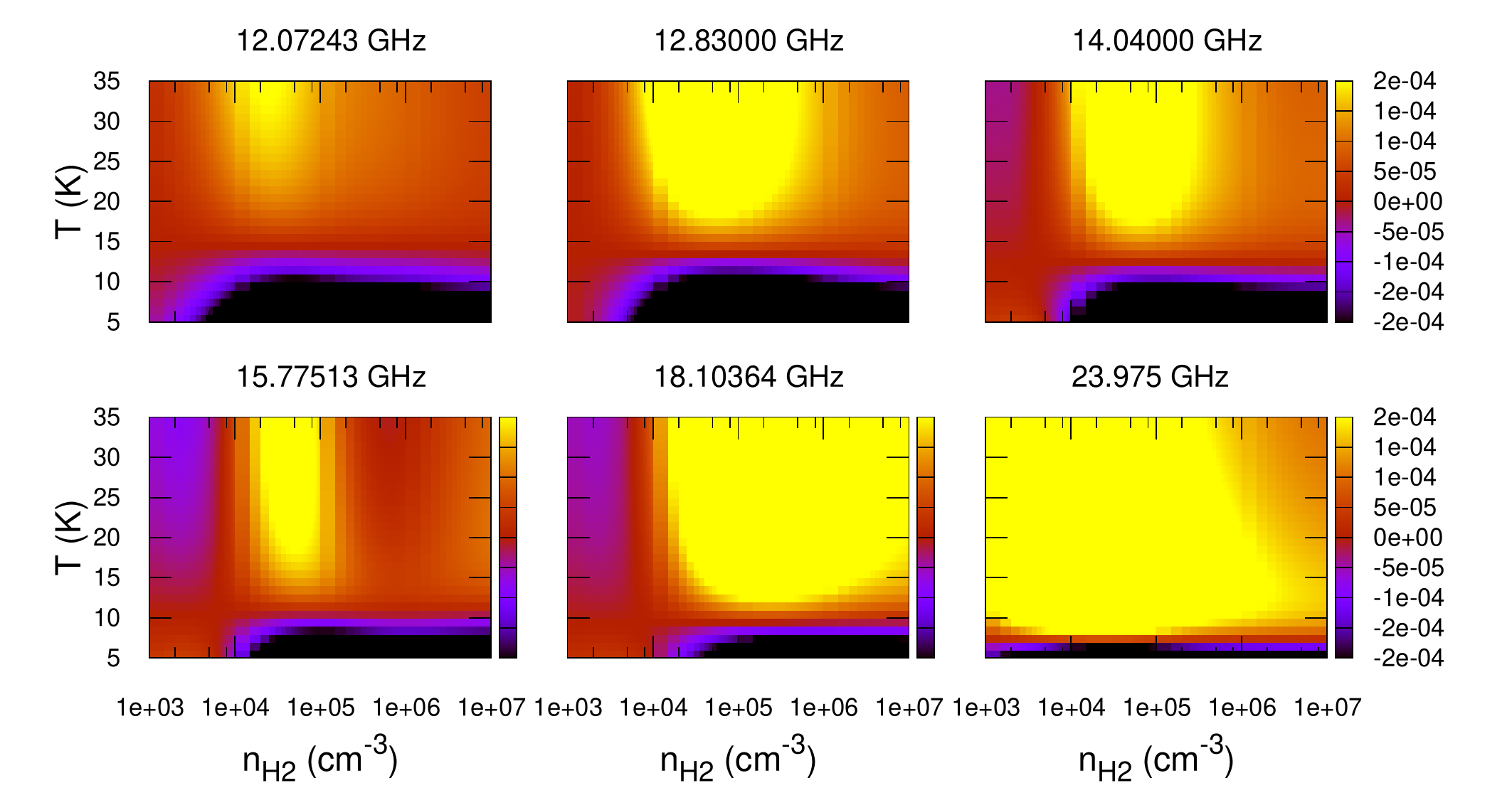}
\caption{Parameter space for the radiation temperature of the most probable $6$ transitions with non-LTE condition.}
\end{figure}

We have further worked on these $6$ transitions to find out the region where it is showing the
absorption feature. In this effort, we have used a wide range of parameter space (by varying the
number density and kinetic temperature of the medium)
to find out the probable absorption zone for these transitions.
We have used the RADEX program again for this purpose. In Fig. 8, we have shown the variation of the radiation temperature depending
on the $\rm{H_2}$ density ($10^4-10^7$ cm$^{-3}$) and kinetic temperature ($5-35$ K). It depicts that the
absorptions are prominent around the low kinetic temperature ($5-10$ K) and for $n_{H2}> 10^4$ cm$^{-3}$. Value of the
color box at the end of the panels represents the obtained radiation temperature of the transitions in K.

In addition to the $3$ transition of PrO discussed above, \cite{mcgu16} observed $18$ transitions of 
acetone and $11$ transitions of propanal from PRIMOS survey.
To test the fate of these transitions with the collisional rate file adopted from e-CH$_3$OH, we have calculated the 
radiation temperature of these transitions for a wide range of parameter space.
We have further explored the validity of our collisional data file by checking the computed radiation temperature of
propanal and acetone in PRIMOS. In Fig. A1 and Fig. A2 (see Appendix), we have shown the radiation temperature of $11$ transitions
of propanal and $18$ transitions of acetone which were observed in PRIMOS.
Fig. A1 shows that for the first six transitions of propanal, we have obtained the absorption feature at $T_{ex}=5-7$ K and
relatively at higher densities ($n_{H2}>10^5$ cm$^{-3}$). 
For the transitions at $29.619$ GHz and $31.002$ GHz, we have obtained the absorption features at
relatively lower $\rm{H_2}$ density ($10^3-10^4$ cm$^{-3}$) and kinetic temperature ($10-35$ K). 
We have noticed that the last three transitions of propanal are showing emission for the whole range of density and temperature
adopted here.
Similarly in case of Fig. A2, for the first $14$ transitions of acetone,
we have obtained the absorption feature within a minimal zone of parameter space, but for the last $4$ transitions
we have not received the absorption feature. 
Since then we have seen a significant discrepancy between our calculated and observed
radiation temperature at the higher frequencies, it 
could be attributed due to the unavailability of the collisional rates and its present approximation.\\

\begin{figure}
\centering{
\includegraphics[height=8cm, width=9cm]{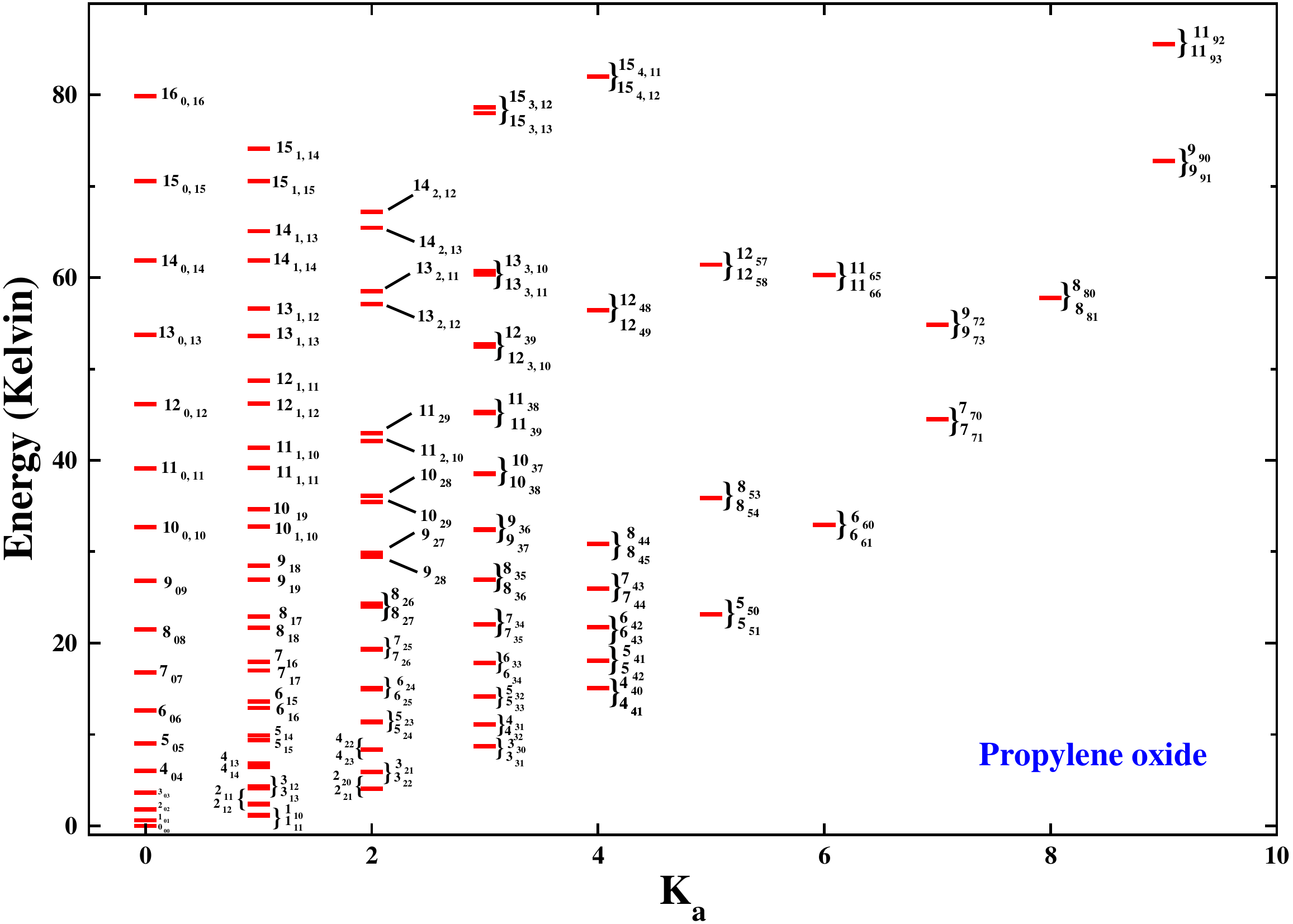}
\caption{Rotational energy level diagram of propylene oxide.}}
\end{figure}

\begin{figure}
\centering{
\includegraphics[height=8cm, width=9cm]{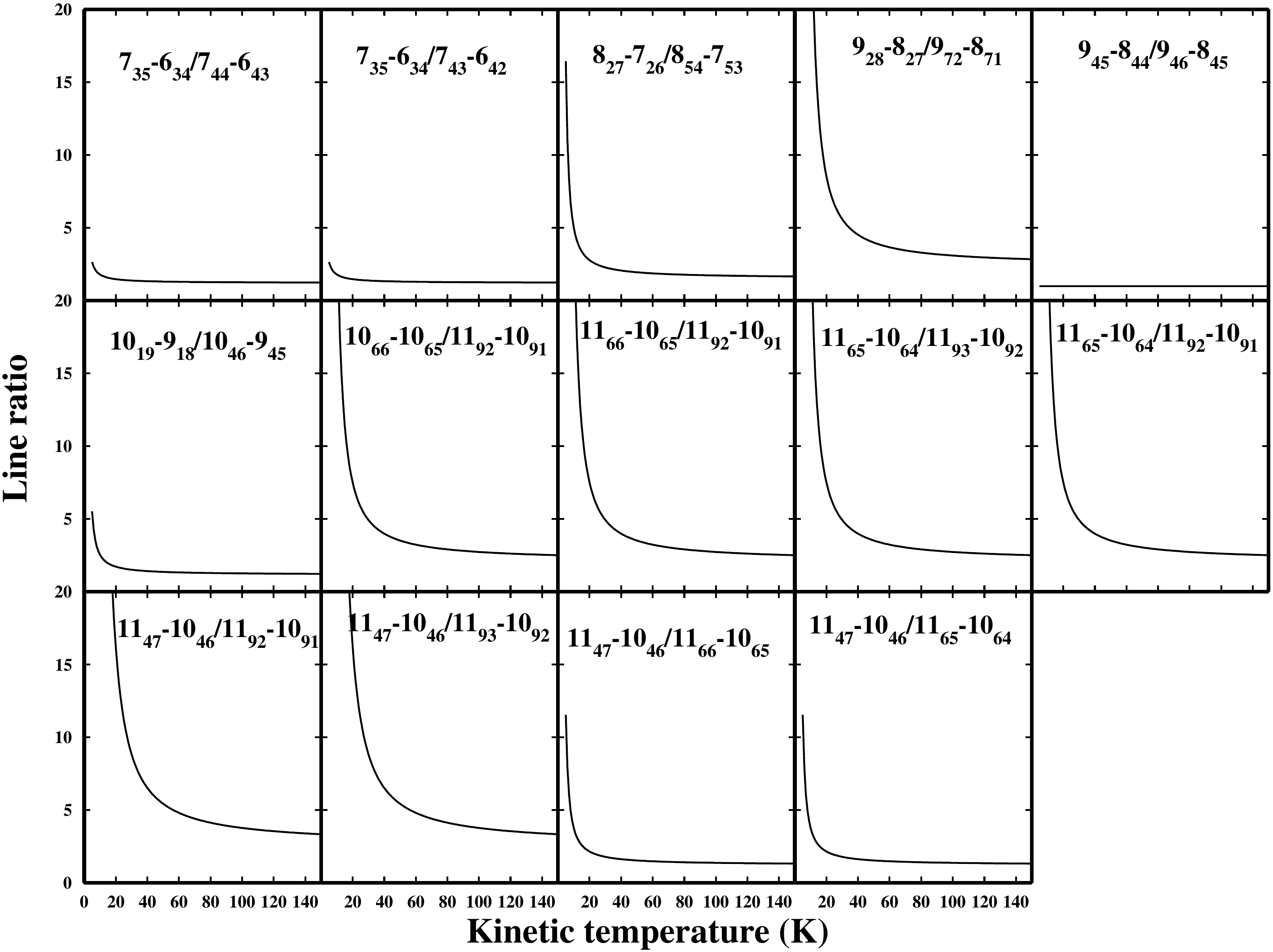}
\caption{Variation of line ratio with the kinetic temperature in LTE approximation.}}
\end{figure}

Our non-LTE model accurately reproduces the three observed transitions of PrO in absorption, and for this, we have used
observed parameters such as excitation temperature, FWHM, and column density of PrO following \cite{mcgu16}. 
We have also found a good agreement between the observed transitions of propanal and acetone with our calculated non-LTE
transitions. Parameter space for the radiation temperature of the most probable transitions of PrO, propanal, and acetone are presented (Fig. 8, Fig. A1 \& Fig. A2) with the non-LTE condition. The similar results obtained by observation \citep{mcgu16} and radiative transfer model are 
used to constrain the physical conditions (density, temperature) of that region where these molecules were observed.\\

In Table 9, we already have pointed out the line parameters of various PrO transitions for the hot core region which lies in ALMA Band 3 and Band 4 regions. For all the transitions shown in Table
9. we have shown the variation of radiation temperature in Fig. A3 for a wide range of parameter space. 
In constructing the parameter space for the hot core region, we have used
$T_{ex}=5-160$ K and $n_{H2}=10^4-10^7$ cm$^{-3}$. Outside each panel, a color box is placed to show the radiation
temperature in K. Fig. A3 would be very useful to know the emission feature of these viable transitions before its
real observation.\\

\subsection{Estimation of the physical properties from the obtained line parameters:}
\cite{mang93} used formaldehyde (H$_2$CO) molecule to trace kinetic temperature
and spatial density within molecular clouds. Normally, symmetric rotor molecules ($\rm{NH_3}$, $\rm{CH_3CN}$, $\rm{CH_3C_2H}$ etc.) are 
used for the measurements of kinetic temperature. However, due to the spatially variable abundances as well as complex
excitation properties, usage of these species are doubtful.
The main reason behind the choice of formaldehyde was that it permeates the ISM at a
relatively high abundance. Formaldehyde is a slightly asymmetric rotor molecule which should
possess closer kinetic temperature sensitivities like purely symmetric rotors.
Since PrO is an asymmetric top molecule, the ratio of lines from different J-states would be used as the density tracers and the ratio of
lines from the same J-state but different K-states could be used as the probes of the regional temperature.\\

\subsubsection{Kinetic temperature measurements:}
In the case of symmetric
rotors, $\Delta K_a=0$ dictates the dipole selection rules. Transitions between other $K_a$-values are
only possible by the collisional excitation. Due to this unique feature, a comparison
between the energy level populations from different $K_a$ levels
within the same symmetry species may be used as a measure of the kinetic temperature. If we consider the
ratio between two transitions which could be measured within the same frequency band, various observational
uncertainties could be nullified.

Propylene oxide is an asymmetric rotor. We can check the line ratio of propylene oxide for the measurement of temperature
and check the consistency of this method with that obtained for H$_2$CO.
Figure 9 depicts the rotational energy level diagram
of propylene oxide.
Because of its complexity, PrO is not as widespread as H$_2$CO.
On the other hand, PrO is also a slightly asymmetric top molecule, and thus we can compare the line ratios obtained by
the PrO observation with the H$_2$CO observation to derive the precise kinetic temperature of the source.
The collisions primarily connect different $K_a$ ladders of PrO. Thus the relative population
occupying the ground states of these two
$K_a$ ladders are related by the Boltzmann equation at Kinetic temperature.
Based on the selection process of \cite{mang93} for the measurement of the kinetic temperature, 
we have used the following selection rule to have the line ratio between two transitions. 
Suppose the line ratio (R) between the two transitions is:
$$
R= \frac{J1 K_{a1} K_{c1} - J2 K_{a2} K_{c2}}{ J3 K_{a3} K_{c3}- J4 K_{a4} K_{c4}},
$$
where, $Ji$, $K_{ai}$, $K_{ci}$ $(i=1,4)$ are used to denote the quantum numbers of the energy levels involved in the 
transition. We have utilized the following selection rules for the measurement of the kinetic temperature from the line ratio.
 (a) $\Delta J = 1$ (i.e., $J1-J2=J3-J4=1$),
 (b) $J1=J3$ and $J2=J4$,
(c)  $\Delta K_a = 0$ (i.e., $K_{a1}-K_{a2}=K_{a3}-K_{a4}=0$),
(d) $K_{a1} \ne K_{a3}$, $K_{a1} \ne K_{a4}$, $K_{a2} \ne K_{a3}$, $K_{a2} \ne K_{a4}$,
(e)  $\Delta K_c = 1$ (i.e., $K_{c1}-K_{c2}=K_{c3}=K_{c4}=1$),
 (f) frequency should be closely spaced. 
Unfortunately, the observed three transitions \citep{mcgu16} did not fulfill the above criterion. Thus, we are
unable to predict the kinetic temperature from the line ratio method here. Alternatively, here, we have focused
on the transitions of PrO which might be observed in the hot core region and identified
some of the transitions which might be useful to find out the physical properties of the perceived source.
In Table 9, we already have selected some transitions based on the obtained intensity and
avoiding blending with possible interstellar species. Among them, we find out some transitions for the
measurement of the kinetic temperature.
The advantage of plotting these line ratios are that these are less sensitive to the calibration
errors and both the lines would have been equally affected.
It is worthy to calculate several transition ratios to avoid the discrepancy on measuring the kinetic temperature.
We have found out following line pairs for the measurement of the kinetic temperature:
$7_{35}-6_{34}/7_{44}-6_{43}(E_{7_{44}}=25.98 K),
7_{35}-6_{34}/7_{43}-6_{42}(E_{7_{43}}=25.98 K),
8_{27}-7_{26}/8_{54}-7_{53}(E_{8_{54}}=35.89 K),
9_{28}-8_{27}/9_{72}-8_{71}(E_{9_{72}}=54.83 K),
9_{45}-8_{44}/9_{46}-8_{45}(E_{9_{46}}=36.32 K),
10_{19}-9_{18}/10_{46}-9_{45}(E_{10_{46}}=42.40 K),
10_{66}-10_{65}/11_{92}-10_{91}(E_{11_{92}}=85.53 K),
11_{66}-10_{65}/11_{92}-10_{91}(E_{11_{92}}=85.53 K),
11_{65}-10_{64}/11_{93}-10_{92} (E_{11_{93}}=85.53 K),
11_{65}-10_{64}/11_{92}-10_{91} (E_{11_{92}}=85.53 K),
11_{47}-10_{46}/11_{92}-10_{91} (E_{11_{93}}=85.53 K),
11_{47}-10_{46}/11_{93}-10_{92} (E_{11_{93}}=85.53 K),
11_{47}-10_{46}/11_{66}-10_{65} (E_{11_{66}}=60.28 K),
11_{47}-10_{46}/11_{65}-10_{64} (E_{11_{65}}=60.28 K)$.
The line ratios of PrO mentioned here have the transitions from various $K_a$ ladders (i.e., with the
$K_a=3,4,5,6,7,8,9$). It is obvious that around the low-temperature region, the ground state of the $K_a$ 
ladders are well populated and lower $K_a$ values would pile up earlier than that of the higher $K_a$ values.\\

From Fig. 9, the ground state of $K_a=3,4,5,6,7,8$ and $9$ ladder is $3_{31}$, $4_{41}$, $5_{51}$, $6_{61}$, $7_{71}$, $8_{81}$
and $9_{91}$ respectively. The upstate energy of the each transitions considered here are all below $100$ K.
Let us first check under LTE approximation how these ratios can behave. At the high densities, LTE conditions are best suited.
Therefore, the relative population occupying the ground states from the two $K_a$ ladders should obey the Boltzmann equation
at the kinetic temperature.
\cite{mang93} derived the following relation for the estimation of kinetic temperature from the obtained line ratio if both
the transitions are optically thin:
\begin{equation}
T_K=[E(J,K')-E(J,K)][ln(\frac{S_{JK'}\int J(T_R(J,K)d\nu}{S_{JK}\int J(T_R(J,K')d\nu})]^{-1},
\end{equation}
where, $E(J,K')$ is the upstate energy of the bottom transition, E(J, K) is the upstate energy of the transition at the numerator,
$S_{JK'}$ and $S_{JK}$ are the transition line strengths of the two upstate involved in the transition ratio and $T_R(J,K')$,
$T_R(J, K)$ are the obtained radiation temperatures. By using the above relation and using the line strengths from
\url{https://www.cv.nrao.edu/php/splat/advanced.php}, we have
computed the line ratio of the radiation temperature for the above transitions.
Fig. 10 shows the variation of these line ratio with the different kinetic temperature. It is interesting to note that around
the low temperature, this ratio is maximum and drastically reduced with the increasing temperature. After a certain temperature,
it remains roughly invariant (small decreasing slope) on the increase of the kinetic temperature.
This feature would be better understood if we explain it in terms of the 1$^{st}$ transition ratio,
$7_{35}-6_{34}/7_{44}-6_{43}(E_{7_{44}}=25.98 K)$ (see first panel of Fig. A4).
Around the low temperature ($T_K<25.98$ K) the ground state of the $K_a=3$
(i.e., $3_{31}$) and $K_a=4$ ladder (i.e., $4_{41}$) are mostly populated.
Since $6_{34}$ level is lying below the $6_{43}$ level, its population is comparatively higher at the low temperature.
Similarly, $7_{35}$ level lies below the $7_{44}$ level implies that the population is higher in $7_{35}$ level.
As we have increased the temperature, population
at the ground state falls much faster.
Since the upper and lower level of the transition at the denominator is at a relatively higher energy level than that of the upper
and lower energy level of the transition at the top,
population in the $6_{43}$ level falls much faster than $7_{44}$ level and
show faster-increasing trend of $7_{44}-6_{43}$ transition in comparison to the top transition.
At $T_K > E_u$ the decreasing slope of the ratio changes and at $T_K >> E_u$ it tends to become roughly
invariant on the kinetic temperature.
The expression provided in eqn. 3 is independent of the
the number density of the cloud and thus providing the kinetic temperature at the LTE condition.
Since, we have considered the hot core region, where, the $\rm{H_2}$ density should be on the higher side (may vary in between $\sim 10^5- 10^7$ cm$^{-3}$) 
and temperature in between $100-200$ K, LTE condition is best suited. Since from various observations, we know the kinetic temperature
of Sgr B2 \citep{bonf17}, we may constrain the line ratio of PrO around this region. Our obtained line ratios from the LTE condition
are presented in Table 10 at $T_K=150$ K.\\

Since we already have prepared our approximated collisional rate file, we have
again calculated these line ratios with the non-LTE calculation for a wide range
of parameter space.
In our calculations, we have varied the kinetic temperature from
$5$ to $160$ K and kept the column density of PrO
constant at $10^{13}$ cm$^{-2}$. We also have considered the column density as high as $1.74 \times 10^{16}$ cm$^{-2}$
(obtained from our two-phase model presented in Fig. 4 and column density is obtained by assuming a $H_2$ column
density of $10^{24}$ cm$^{-2}$)
to see the effect on the calculated ratio but did not find any drastic difference.
In Fig. A4, we have shown the line ratios which might
be useful to precisely trace the temperature of the
region where these lines would be observed. In the top of the panel, we have mentioned the
upstate energy of four energy states which were involved in the transition ratio.
In Table 10, we have also noted down our line ratios obtained from the non-LTE calculations with 
$n_H=10^7$ cm$^{-3}$ and $T_{ex}=150$ K.
It should be noted that the line ratios obtained from the LTE calculation are uncertain by $<30\%$ \citep{mang93} and
we have adopted random approximation for the consideration of our collisional data file.
Despite this fact, we have obtained an excellent agreement between the LTE consideration and non-LTE calculation (Table 10).

\subsubsection{Spatial density measurements:}
Measurement of the spatial density is more critical than the measurement of the kinetic temperature.
Based on the study by \cite{mang93}, spatial density could be measured by measuring a specific line ratio.
\cite{mang93} used the line ratio
of various $\rm{H_2CO}$ transitions for the measurement of the spatial density. Here, we have applied
a similar technique to find out the line ratio for some of the observable transitions of
PrO in the hot core region like Sgr B2(LMH).
We have used the following criterion for the selection of the transition ratios:
(a)  $\Delta J = 1$ (i.e., $J1-J2=J3-J4=1$),
(b) $J1 \ne J3$ and $J2 \ne J4$,
(c) $\Delta K_a = 0$ (i.e., $K_{a1}-K_{a2}=K_{a3}-K_{a4}=0$),
(d)  $K_{a1}=K_{a2}=K_{a3}=K_{a4}$,
(e) $\Delta K_c = 1$ (i.e., $K_{c1}-K_{c2}=K_{c3}=K_{c4}=1$),
(f)  frequency should be closely spaced. 
Very rarely the above criteria are satisfied, and sometimes the frequency lies very far away, and thus absolute calibration of each transition is 
required. The observed three transitions \citep{mcgu16} are not satisfying the above criterion. We also did not find any other transitions which 
can satisfy the above criterion around the region (dark cloud condition) where these three transitions were observed. Thus, we have selected some 
more transitions (
$7_{44}-6_{43}/9_{46}-8_{45},
7_{44}-6_{43}/9_{45}-8_{44},
7_{44}-6_{42}/9_{46}-8_{45},
7_{43}-6_{42}/9_{45}-8_{44},
8_{27}-7_{26}/9_{28}-8_{27}$)
 from the LTE model of the hot core region (Table 9) which can 
fulfill the above criterion and suitable for the observation in the hot core region.
To constrain a limit on the number density, in Fig. A5, we have shown the line ratios of the
PrO transitions for $T_{ex}=150$ K. Here, the parameter space is based on the column density of PrO and the
$\rm{H_2}$ number density of the medium. Column density have been varied in between $10^{13}$ cm$^{-2}-10^{16}$ cm$^{-2}$ 
and $\rm{H_2}$ number density of the medium has been varied in between $10^4-10^7$ cm$^{-3}$.
The density and temperature of the Sgr B2 (LMH) region were well studied, and it would be assumed that the 
$\rm{H_2}$ density may vary in between $10^6-10^7$ 
cm$^{-3}$ and kinetic temperature may vary in between $100-150$ K. 
Assuming the predicted upper limit of the PrO column density of $6.7 \times 10^{14}$ cm$^{-2}$ from \cite{cunn07}, we have the
line ratio of $0.55,\ 0.57,\ 0.57, \ 0.59$ and $0.8$ for the above 5 transitions respectively in the Sgr B2(LMH).

\renewcommand{\arraystretch}{0.7}
\begin{table}
{\scriptsize
\centering
\scalebox{0.001}
\tiny{ 
\caption{LTE line parameters of the various transitions of PrO using ALMA.}
\begin{tabular}{|p{0.7 in}|p{0.4 in}|p{0.2 in}|p{0.35 in}|p{0.4 in}|p{0.1 in}|p {0.3 in}|p {0.4 in}|}
\hline
Transition ($J_{Ka'kc'}-J_{ka''kc''}$) &Frequency (MHz)& E$_{up}$ (K) & A$_{ij}$ (s$^{-1}$) & Tau & Tex (K) & Intensity (K)\\
\hline
\hline
16 2 15 - 16 1 16  & 85166.334  &  83.96 & 2.4E-06 &  1.30E-02 & 150  & 0.58\\
7  1  7 - 6  1  6  & 85484.159  &  17.01 & 3.0E-06  &  1.13E-02 & 150  & 0.51\\
17 4 14 - 17 3 15  & 85856.007  & 102.12 & 5.5E-06  &  2.73E-02 & 150  & 1.23\\
4  2  2 - 3  1  3 & 88348.483$^a$ &   8.33  &  3.8E-06 &  8.63E-03 & 150  & 0.39\\
7  3  5 - 6  3  4 & 88654.935  &  22.06 & 2.8E-06  &  9.49E-03 & 150  & 0.43\\
7  4  4 - 6  4 3  & 88599.017  &  25.98 & 2.3E-06  &  7.62E-03 & 150  & 0.35\\
7  4  3 - 6 4 2  & 88601.917  &  25.98 & 2.3E-06   &  7.62E-03 & 150  & 0.35\\
 7  1  6 - 6  1  5 & 90475.090 &   17.98 &  3.5E-06 &  1.19E-02 & 150  & 0.55\\
 17 2 16 - 17 1 17 & 90723.178 &   94.08 &2.8E-06   &  1.30E-02 & 150  & 0.65\\
 7  1  7 - 6 0 6   & 91337.880$^{a,b}$ &   17.01 &  8.0E-06 &  2.65E-02 & 150  & 1.24\\
 8 0  8 - 7  1 7   & 94045.665$^a$&   21.52 &  8.7E-06 &  3.00E-02 & 150  & 1.43\\
 3  3  1 - 2  2  0 & 96421.008$^{a,b}$ &    8.70 &  1.0E-05 &  1.52E-02 & 150  & 0.73\\
 18 5 13 - 18 4 14 & 97912.135 &  118.10 & 8.4E-06  &  3.04E-02 & 150  & 1.49\\
 8  0  8 - 7  0  7 & 98662.010 &    21.52&  4.7E-06 &  1.47E-02 & 150  & 0.72 \\
 8  2  7 - 7  2  6 & 100647.469&    24.03&  4.7E-06 &  1.40E-02 & 150  & 0.70\\
 8  1  8 - 7  0  7 & 102188.761$^a$&    21.69&   1.2E-06&  3.41E-02 & 150  & 1.70\\
 5  2  3 - 4  1  4 & 103094.044&    11.39&   4.9E-06&  9.71E-03 & 150  & 0.49\\
 14 5  9 - 14 4  10& 103204.976&    77.85&  8.9E-06 &  2.99E-02 & 150  & 1.51\\
 6  2  5 - 5  1  4 & 105436.485$^a$&    14.97&   7.2E-06&  1.57E-02 & 150  & 0.81\\
 9  0  9 - 8  1  8 & 107006.105$^{a,b}$&    26.82&   1.4E-05&  3.93E-02 & 150  & 2.01\\
 4  3  2 - 3  2  1 & 108983.168$^a$&    11.12&   1.2E-05&  1.75E-02 & 150  & 0.91\\
 4  3  1 - 3  2  2 & 109160.580$^a$&    11.12&   1.2E-05&  1.74E-02 & 150  & 0.91\\
 9  2  8 - 8  2  7 & 113085.512&    29.46&  6.8E-06 &  1.73E-02 & 150  & 0.91\\
 9  1  9 - 8  0  8 & 113153.015$^{a,b}$&    26.95&   1.7E-06&  4.26E-02 & 150  & 2.24\\
 9  7  2 - 8  7  1 & 113828.145&    54.83&   2.9E-06&  6.12E-03 & 150  & 0.32\\
 9  4  6 - 8  4  5 & 114012.677&    36.32&   5.9E-06&  1.41E-02 & 150  & 0.74\\
 9  4  5 - 8  4  4 & 114031.614&    36.32&   5.9E-06&  1.41E-02 & 150  & 0.74\\
\hline
\hline
 17 6 11- 17 5 12 & 126748.398&   113.19&   1.6E-05&   3.50E-02&  150&  1.94\\
 10 4 6 - 9 4 5   & 126783.156&    42.40&   8.6E-06&   1.75E-02&  150&  0.97\\
 14 6 8 - 14 5 9  & 127852.183&    83.99&   1.6E-05&   3.30E-02&  150&  1.83\\
 15 6 10 - 15 5 11& 127659.038&    93.11&   1.6E-05&   3.41E-02&  150&  1.89\\
 16 6 11 - 16 5 12& 127377.638&   102.84&   1.6E-05&   3.48E-02&  150&  1.93\\
 9 6 3 - 9 5 4    & 128562.724&    47.53&   1.2E-05&   2.10E-02&  150&  1.17\\
 10 6 5 - 10 5 6  & 128487.914&    53.60&   1.3E-05&   2.44E-02&  150&  1.36\\
 13 6 7 - 13 5 8  & 128075.357&    75.48&   1.5E-05&   3.16E-02&  150&  1.75\\
 13 6 8 - 13 5 9  & 128097.570&    75.48&   1.5E-05&   3.16E-02&  150&  1.75\\
 12 6 6 - 12 5 7  & 128249.483&    67.58&   1.5E-05&   2.96E-02&  150&  1.65\\
 10 1 9 -9 1 8    & 128225.980&    34.64&   1.0E-05&   2.19E-02&  150&  1.22\\
 12 6 7 - 12 5 8  & 128259.465&    67.58&   1.5E-05&   2.96E-02&  150&  1.65\\
 10 6 4 - 10 5 5  & 128486.347&    53.60&   1.3E-05&   2.44E-02&  150&  1.36\\
 11 0 11 - 10 1 10& 132290.717&    39.14&   2.8E-05&   5.94E-02&  150&  3.32\\
 6 3 4 -5 2 3     & 133619.742&    17.81&   1.7E-05&   2.26E-02&  150&  1.28\\
 11 9 2 - 10 9 1  & 139104.681&    85.53&   4.5E-06&   6.26E-03&  150&  0.36\\
 11 9 3 - 10 9 2  & 139104.681&    85.53&   4.5E-06&   6.26E-03&  150&  0.36\\
 11 8 3 - 10 8 2  & 139131.647&    75.99&   6.4E-06&   9.50E-03&  150&  0.54\\
 11 8 4 - 10 8 3  & 139131.647&    75.99&   6.4E-06&   9.50E-03&  150&  0.54\\
 11 6 6 - 10 6 5  & 139230.454&    60.28&   9.5E-06&   1.58E-02&  150&  0.91\\
 11 6 5 - 10 6 4  & 139230.502&    60.28&   9.5E-06&   1.58E-02&  150&  0.91\\
 11 4 7 - 10 4 6  & 139567.307&    49.10&   1.2E-05&   2.10E-02&  150&  1.21\\
16 7 10 - 16 6 11 & 151281.705&   110.10&   2.5E-05&   3.69E-02&  150&   2.16\\
14 7 8 -14 6 9    & 151616.943&    91.27&   2.4E-05&   3.45E-02&  150&   2.02\\
13 7 7 - 13 6 8   & 151740.138&    82.76&   2.3E-05&   3.25E-02&  150&   1.91\\
8 7 1 - 8 6 2     & 152054.582&    49.36&   1.2E-05&   1.36E-02&  150&   0.80\\
8 7 2 - 8 6 3     & 152054.583&    49.36&   1.2E-05&   1.36E-02&  150&   0.80\\
12 5 8 - 11 5 7   & 152058.401&    61.42&   1.5E-05&   2.20E-02&  150&   1.29\\
12 5 7 - 11 5 6   & 152064.369&    61.42&   1.5E-05&   2.20E-02&  150&   1.90\\
11 2 10 - 10 1 9  & 155300.752&    42.09&   .25E-04&   3.83E-02&  150&   2.26\\
13 1 13 - 12 0 12 & 158512.466&    53.75&   .51E-04&   8.07E-02&  150&   4.74\\
\hline
\end{tabular}}\\
$^a$ Attempted in Orion KL by \cite{cunn07}\\
$^b$ Attempted in Sgr B2 by \cite{cunn07}\\
ALMA Band 3 = 84-116 GHz\\
ALMA Band 4 = 125-163 GHz\\}
\end{table}

\begin{table}
\scriptsize
\centering
\caption{Estimated line ratio of PrO for the measurement of Kinetic
temperature at $150$K from LTE and non-LTE.}
\begin{tabular}{|c|c|c|}
\hline
Line ratio (upstate energy in K)&LTE results&non-LTE results\\
\hline
$7_{35}-6_{34}/7_{44}-6_{43} (E_{7_{44}}=25.98)$&1.24&1.25\\
$7_{35}-6_{34}/7_{43}-6_{42} (E_{7_{43}}=25.98)$&1.24&1.25\\
$8_{27}-7_{26}/8_{54}-7_{53} (E_{8_{54}}=35.89)$&1.66&1.65\\
$9_{28}-8_{27}/9_{72}-8_{71} (E_{9_{72}}=54.83)$&2.84&2.78\\
$9_{45}-8_{44}/9_{46}-8_{45} (E_{9_{46}}=36.32)$&0.99&1.00\\
$10_{19}-9_{18}/10_{46}-9_{45} (E_{10_{46}}=42.40)$&1.23&1.25\\
$10_{66}-10_{65}/11_{92}-10_{91} (E_{11_{92}}=85.53)$&2.51&2.53\\
$11_{66}-10_{65}/11_{92}-10_{91} (E_{11_{92}}=85.53)$&2.51&2.53\\
$11_{65}-10_{64}/11_{93}-10_{92} (E_{11_{93}}=85.53)$&2.51&2.53\\
$11_{65}-10_{64}/11_{92}-10_{91} (E_{11_{92}}=85.53)$&2.51&2.53\\
$11_{47}-10_{46}/11_{92}-10_{91} (E_{11_{93}}=85.53)$&3.35&3.32\\
$11_{47}-10_{46}/11_{93}-10_{92} (E_{11_{93}}=85.53)$&3.35&3.32\\
$11_{47}-10_{46}/11_{66}-10_{65} (E_{11_{66}}=60.28)$&1.33&1.31\\
$11_{47}-10_{46}/11_{65}-10_{64} (E_{11_{65}}=60.28)$&1.33&1.31\\
\hline
\end{tabular}
\end{table}

\section{Conclusions}
Presence of chiral species in the ISM been recently confirmed by \cite{mcgu16}. 
Observation of more chiral species and identification of their chirality may provide some
lights on the homochirality of various pre-biotic species.
In this paper, we have studied the chemical evolution of PrO under various physical conditions
and find out the line parameters of its transitions which are relevant to constrain the physical properties of
the source. The major conclusions are the following:

$\bullet$ A complete reaction network for the formation of PrO was prepared. This pathway is implemented
in our gas-grain chemical model to find out its chemical evolution. To understand the formation of
PrO in the cold environment as observed by \citep{mcgu16}, we considered a static cloud model, and for the formation of 
PrO in the hot core region, we considered a two-phase model (static isothermal phase followed by a warm-up phase).
We found that the reaction between $\rm{C_3H_6}$ and O($^3$P) with an activation barrier of $40$ K 
is the dominant means for the formation of PrO in the ice phase. This ice phase PrO desorbed to the gas phase by various 
means which could be observed by their rotational transitions. We also considered the formation of PrO 
by O($^1$D) but we in absence of this we also have significant PrO production due to the consideration of 
lower activation barrier for the reaction between $\rm{C_3H_6}$ and O($^3$P).

$\bullet$ We computed various vibrational transitions of PrO along with its protonated form which might be observable 
with the forthcoming JWST facility. 

$\bullet$ To constrain the physical properties of the cloud, we carried out non-LTE modeling (by considering
approximated collisional data file) for
the most probable transitions identified by the LTE calculations by considering a wide range of parameter space
(spanned by the $\rm{H_2}$ density range of $\sim 10^3-10^7$ cm$^{-3}$ of the collisional partner and the 
kinetic temperature
temperature range of $\sim 5-35$ K).
We found that the observed three transitions at $12.07243$ GHz ($1_{10}-1_{01}$), $12.83734$ GHz
($2_{11}-2_{02}$) and $14.04776$ GHz ($3_{12}-3_{03}$) showed absorption features
 around the high $\rm{H_2}$ density ($10^4-10^7$ cm$^{-3}$) and low temperature ($<10$ K) region. Additionally,
we found three more transitions (at $15.78$, $18.10$ and $23.975$ GHz) which might be observed in absorption around
the same region where the three transitions
were identified by \cite{mcgu16}.

$\bullet$ We identified the line parameters suitable for the hot core observation of PrO. Potentially observable transitions
lying in the ALMA Band 3 and Band 4 are pointed out which would be used for the for future observations.

$\bullet$ Various line ratios were directly found to be linked with the kinetic temperature and spatial density distribution.
Thus from observation, one can extract the physical properties of the star-forming region by comparing their observation with our
proposed parameters.

Our result depends on a very simple-minded cloud model which has no rotation and also the evolution
was done until an estimated time. Thus the estimation of the parameters mentioned above could
slightly depend on the evolution time inside the cloud. These aspects would be
further explored in the future.

\begin{acknowledgements}
This research was possible in part due to a Grant-In-Aid from the Higher Education Department of the Government of 
West Bengal. 
AD is grateful to ISRO Respond project (Grant No. ISRO/RES/2/402/16-17)
PG acknowledge CSIR extended SRF 
fellowship (Grant No. 09/904 (0013) 2018 EMR-I). 
{ We want to acknowledge the reviewer for the numerous suggestion to improve the paper.}

\end{acknowledgements}

%
%

\appendix
\section{Additional Figures}
\clearpage
\vskip 0.1cm

\begin{figure}
\hskip -1.2cm
\centering
\includegraphics[height=10cm,width=10cm,angle=270]{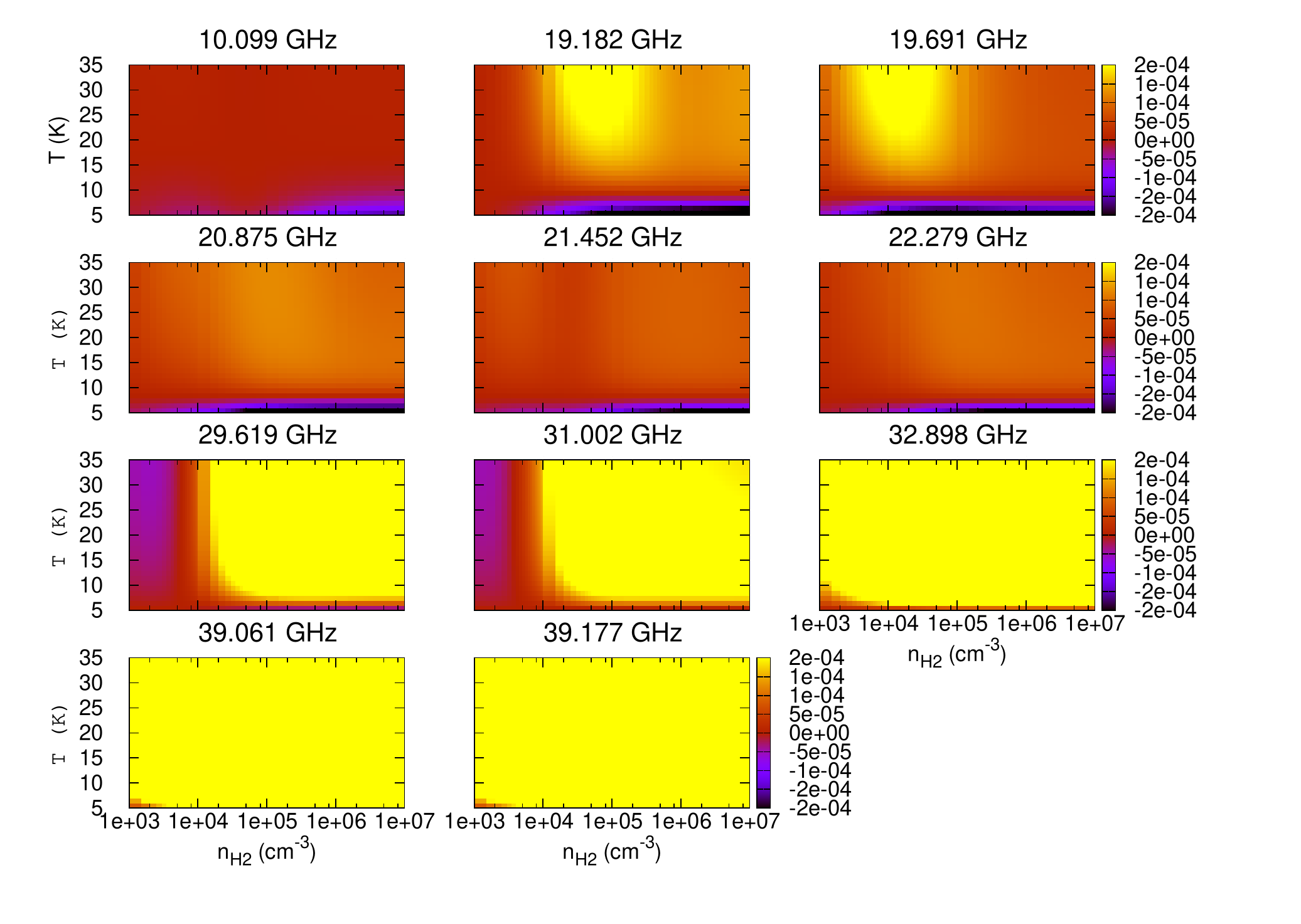}
\vskip 0.1cm
{\small {\bf Fig. A1.} Parameter space for the radiation temperature of the $11$ observed propanal transitions with non-LTE condition.}
\end{figure}

\begin{figure}
\hskip -1.1 cm
\centering
\includegraphics[height=10cm,width=11cm,angle=270]{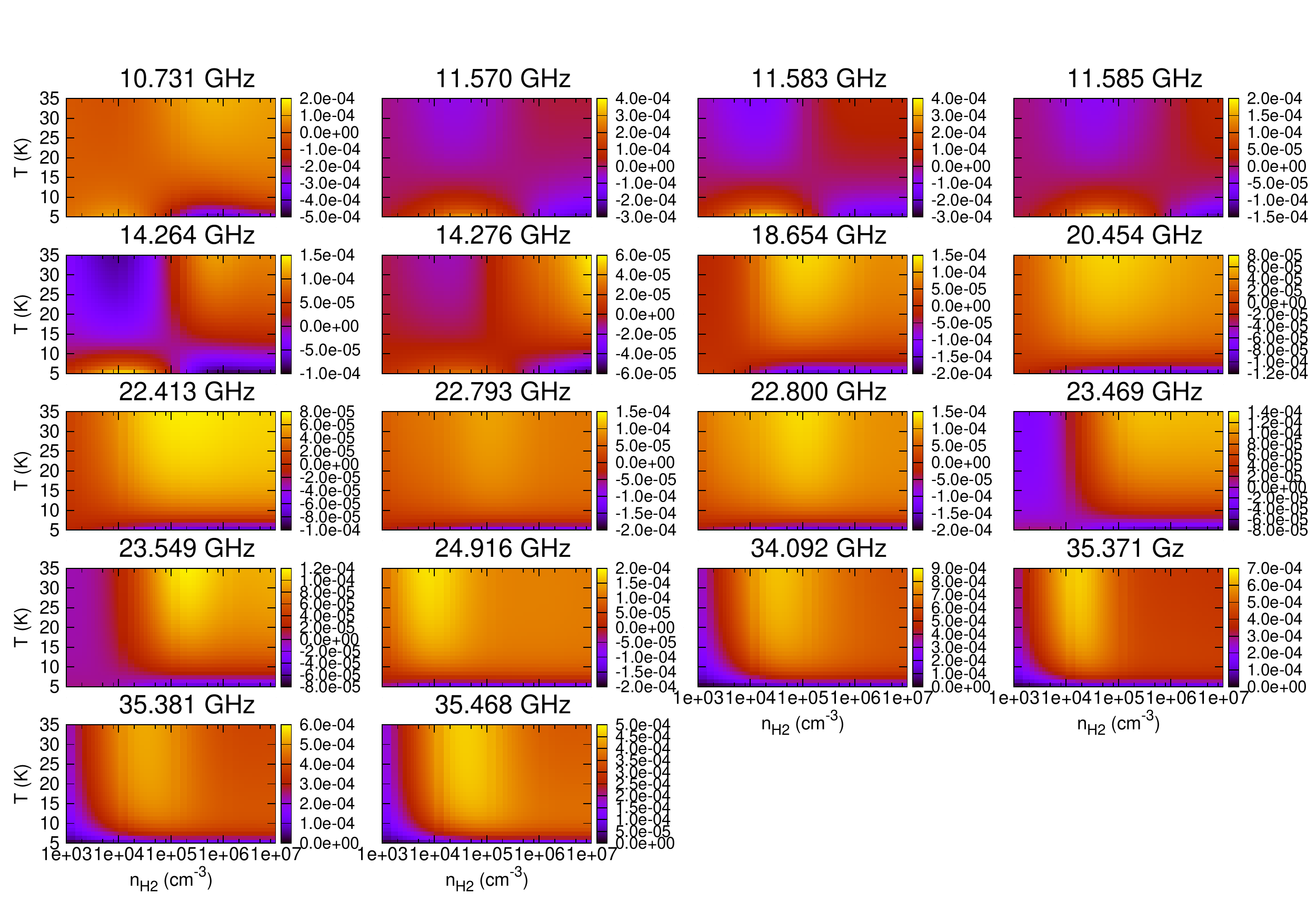}
\vskip 0.1cm
{\small {\bf Fig. A2.} Parameter space for the radiation temperature of the $18$ observed acetone transitions with non-LTE condition.}
\end{figure}

\begin{figure}
\centering
\includegraphics[height=10cm,width=8cm,angle=270]{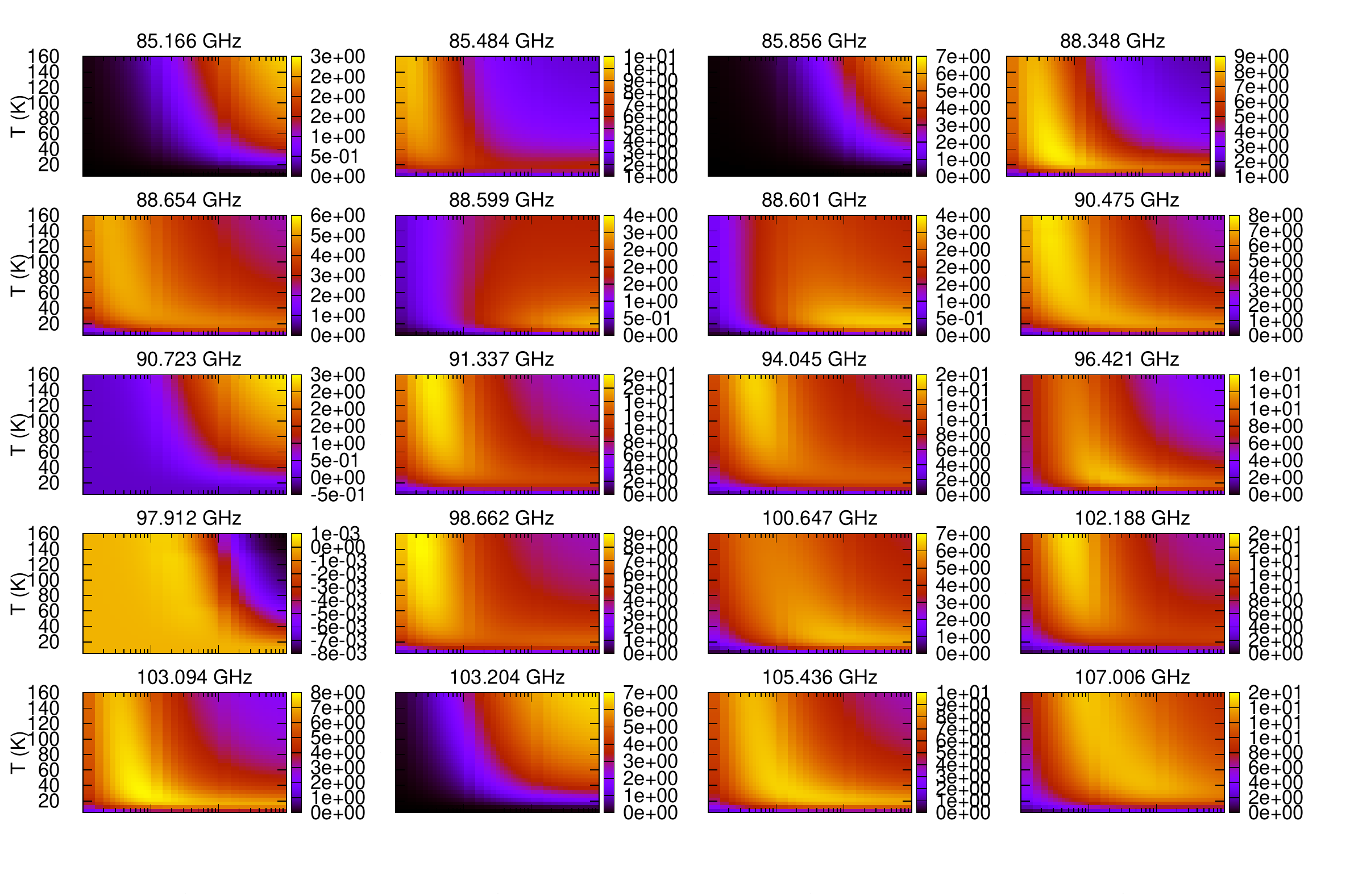}
\vskip -0.89cm
\includegraphics[height=10cm,width=8cm,angle=270]{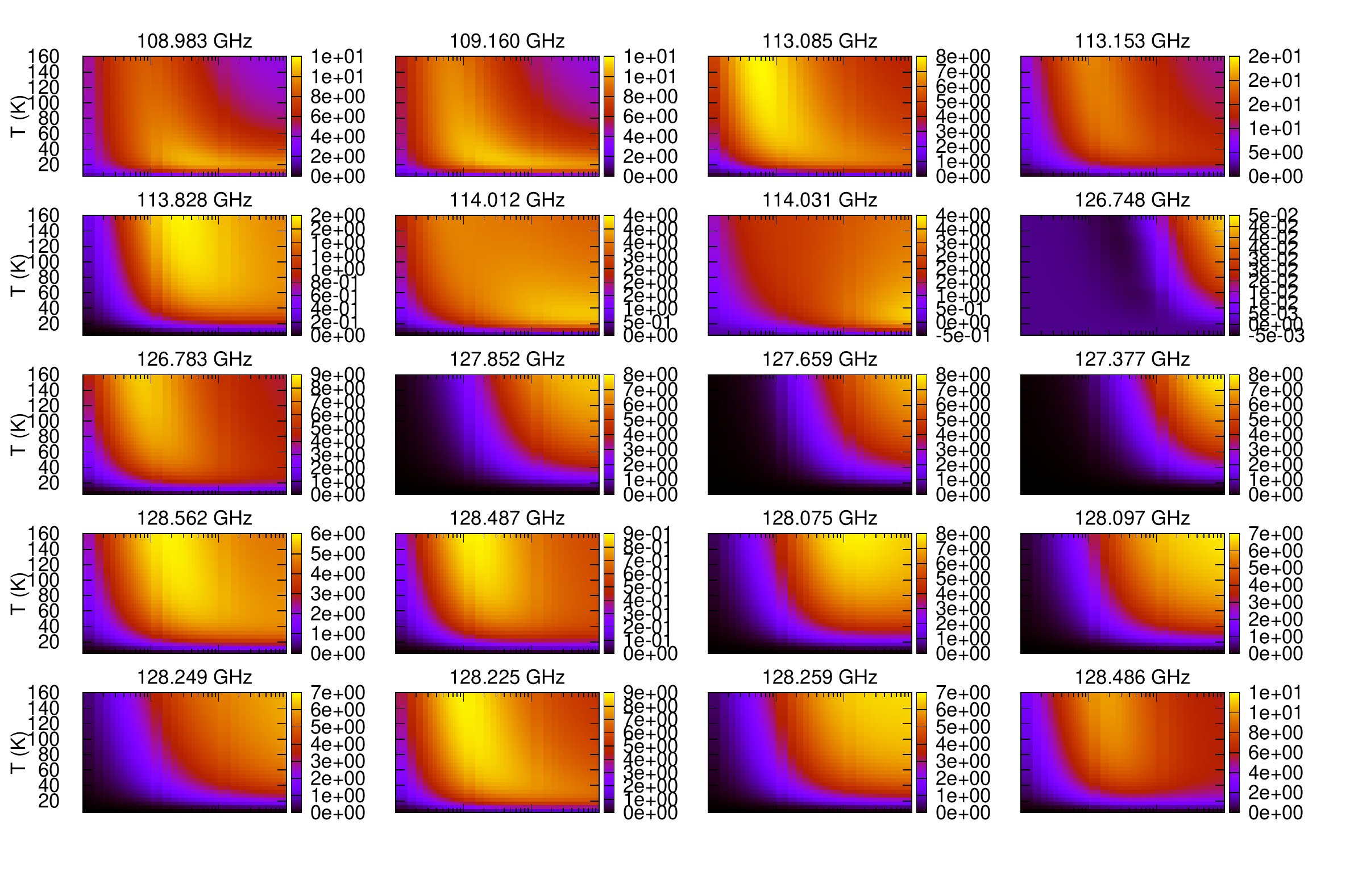}
\vskip -1cm
\includegraphics[height=10cm,width=8cm,angle=270]{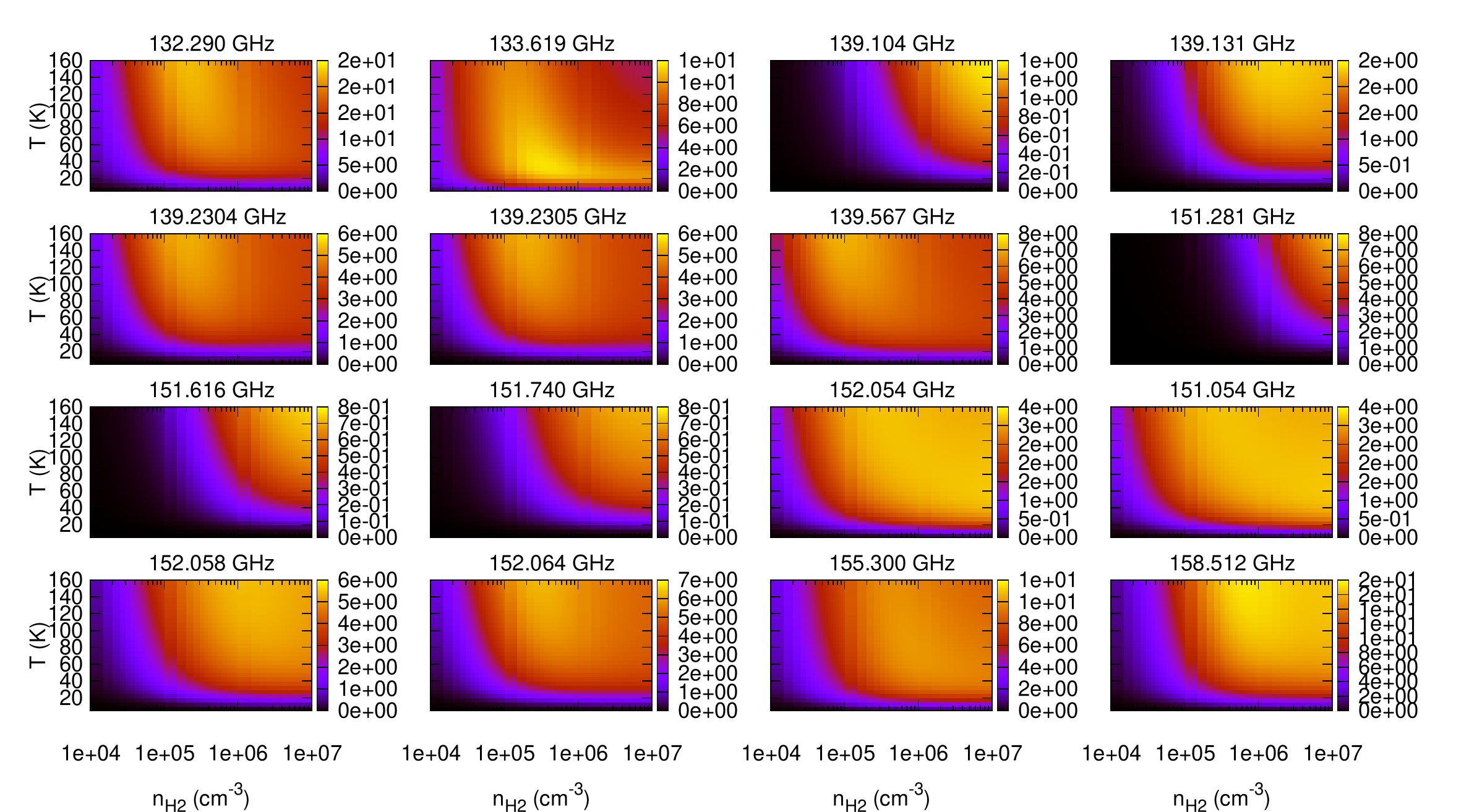}
\vskip 0.1cm
{\small {\bf Fig. A3.} Parameter space of the radiation temperature for the probable PrO transitions in the hot core region.}
\end{figure}

\begin{figure}
\centering{
\includegraphics[height=10cm, width=8cm,angle=-90]{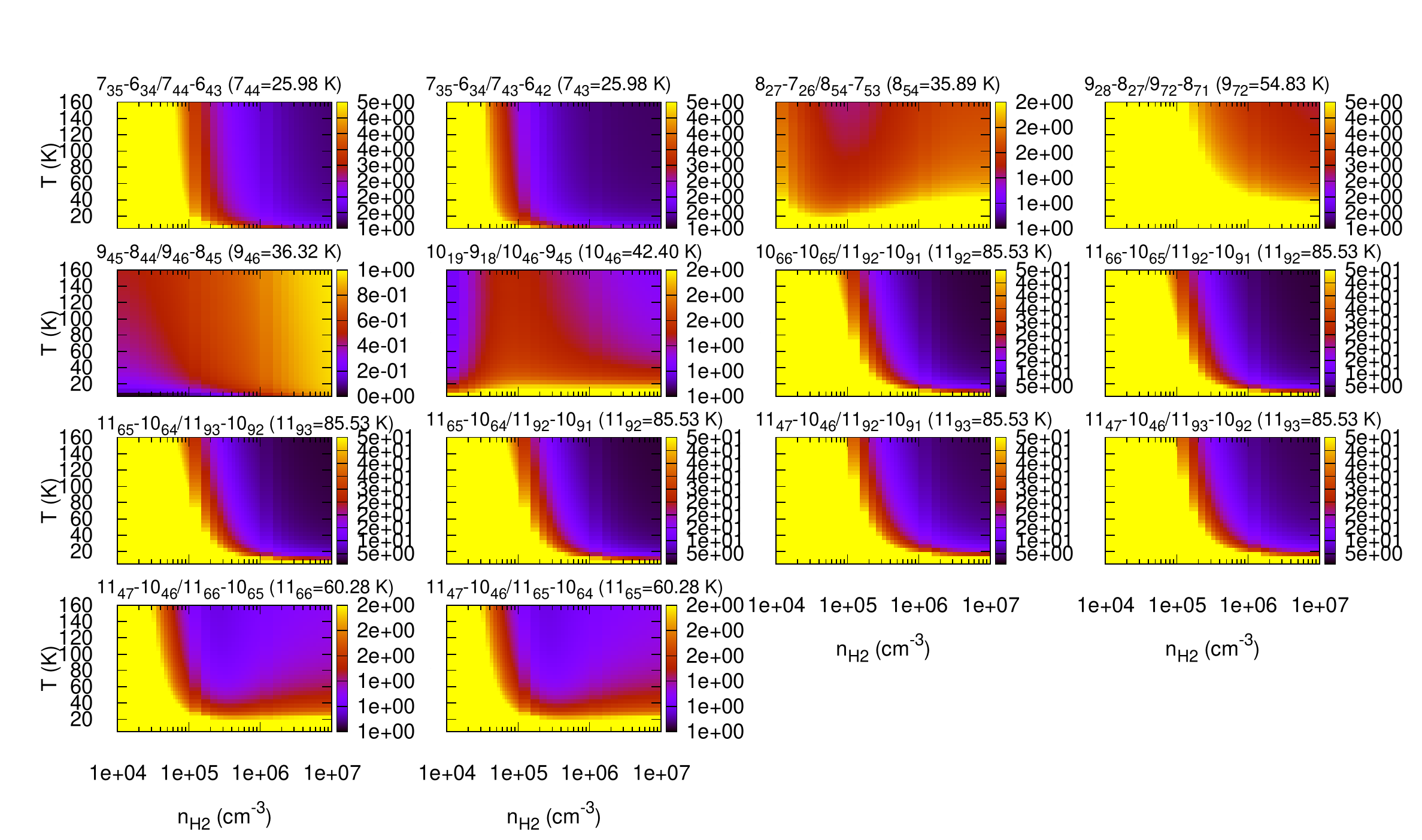}
\vskip 0.1cm
{\small {\bf Fig. A4.} Variation of the line ratios (useful for the kinetic temperature measurement) for a wide range of parameter space (number density and 
{kinetic temperature}).}}
\end{figure}

\begin{figure}
\centering
\includegraphics[height=10cm, width=7cm,angle=-90]{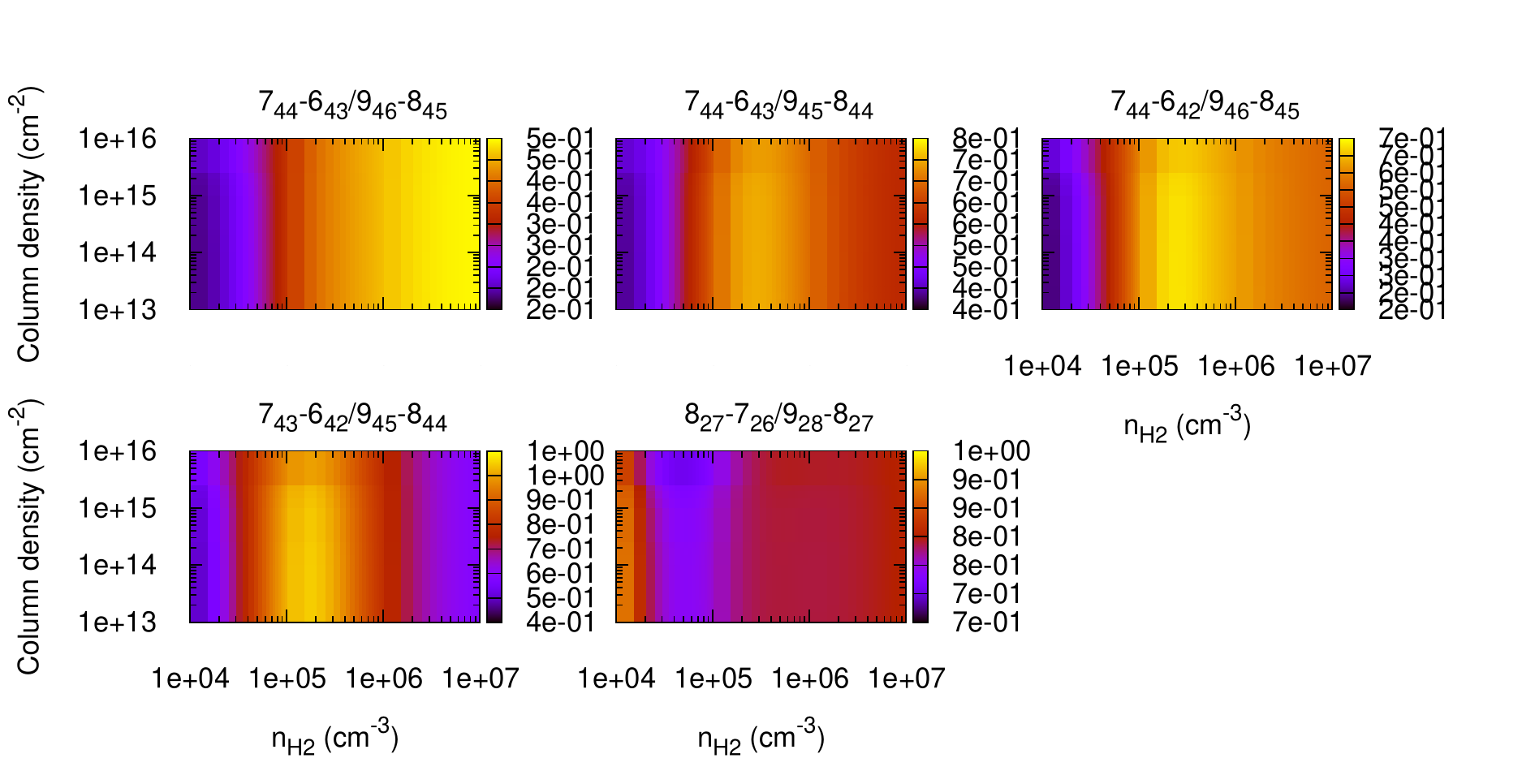}
\vskip 0.1cm
{\small {\bf Fig. A5.} Variation of line ratios (useful for the measurement of the spatial density) for a wide range of parameter space (column density and
number density).}
\end{figure}

\end{document}